\documentclass{aa}  
\usepackage{graphicx}
\graphicspath{{./Figures/}}% ADDED
\usepackage[varg]{txfonts}
\pdfmapfile{+txfonts.map} % ADDED following web info
\usepackage{natbib}
\bibpunct{(}{)}{;}{a}{}{,} % to follow the A&A style
\usepackage{color}% For color text: \color command
\usepackage{amsmath}% for using align
\usepackage{hyperref} % ADDED on A&A suggestion: options to be set?
\begin{document} 
   \title{Dark signal correction for a lukecold frame transfer CCD}
   \subtitle{Application to the SODISM solar telescope onboard the PICARD space mission}
   \author{{J.-F.~Hochedez}\inst{\ref{LATMOS}} \inst{\ref{ROB}}
\and
   {C.~Timmermans}\inst{\ref{ISBA}}
\and
{A.~Hauchecorne}\inst{\ref{LATMOS}}
\and
{M.~Meftah}\inst{\ref{LATMOS}}
}

   \institute{
{{Laboratoire Atmosph\`eres, Milieux, Observations Spatiales} (LATMOS), Centre National de la Recherche Scientifique (CNRS), Universit\'e Paris~VI, Universit\'e de Versailles Saint-Quentin-en-Yvelines, Institut Pierre-Simon Laplace (IPSL), F-78280, Guyancourt, France.}\label{LATMOS}
\email{hochedez@latmos.ipsl.fr}
         \and
{{Observatoire Royal de Belgique -- Koninklijke Sterrenwacht van Belgi\"e} (ORB-KSB), All\'ee Circulaire 3., B-1180 Bruxelles (Uccle), Belgique.}\label{ROB}
\email{hochedez@oma.be}
         \and
{{Institut de Statistique, Biostatistique et Sciences Actuarielles} (ISBA), Voie du Roman Pays 20, L1.04.01, B-1348 Louvain-la-Neuve, Belgique.}\label{ISBA}
             }

   \date{Received Mar.~06., 2013 / Accepted $\star\!\star\!\star\star\!\star\!\star$ } % Month day., Year}

   \abstract
% \abstract{}{}{}{}{} 
% 5 {} token are mandatory
  % context heading (optional)
  % {} leave it empty if necessary  
  % aims heading (mandatory)
  % methods heading (mandatory)
  % results heading (mandatory)
  % conclusions heading (optional), leave it empty if necessary 
{
Astrophysical observations must be corrected for their imperfections of instrumental origin. 
When Charge Coupled Devices (CCDs) are used, their dark signal is one such hindrance.
In their pristine state, most CCD pixels are `cool', \textit{i.e.} they exhibit a low, quasi uniform dark current, which can be estimated and corrected for.
In space, after having been hit by an energetic particle, pixels can turn `hot', \textit{viz.} they start delivering excessive, less predictable, dark current.
The hot pixels need therefore to be flagged so that subsequent analysis may ignore them.
}{
The image data of the PICARD SODISM solar telescope (Meftah et al. 2013) require dark signal correction and hot pixel identification.
Its E2V\,42-80 CCD operates at $-7.2$\degr C and has a frame transfer architecture.
Both image and memory zones thus accumulate dark current during, respectively, integration and readout time.
These two components must be separated in order to estimate the dark signal for any observation.
This is the main purpose of the Dark Signal Model presented in this  paper. 
}{
The dark signal time series of every pixel is processed by the `unbalanced Haar technique' (Fryzlewicz~2007) in order to timestamp the instants when its dark signal is expected to change significantly.
In-between those, both components are assumed constant, and a robust linear regression \textit{vs.} integration time provides first estimates and a quality coefficient.
The latter serves to assign definitive estimates for this pixel and for that period.
}{
Our model is part of the SODISM Level~1 data production scheme.
To check its reliability, we verify on dark frames that it leaves a negligible residual bias (5\,e$^-$), and generates a small RMS error (25\,e$^-$\,rms).
We also analyze the distribution of the image zone dark current.
The cool pixel level is found to be 4.1\,e$^-\!\cdot\mathrm{pxl}^{-1}\!\cdot\mathrm{s}^{-1}$, in agreement with the predicted value.
The emergence rate of hot pixels is investigated too.
It legitimates a threshold criterion at 50\,e$^-\!\cdot\mathrm{pxl}^{-1}\!\cdot\mathrm{s}^{-1}$.
The growth rate is found to be on average $\sim$500 new hot pixels per day, \textit{i.e.} 4.2\% of the image zone area per year.
}{
A new method for dark signal correction of a frame transfer CCD operating at only \textit{ca.} $-10$\degr C is demonstrated.
It allows making recommendations about the scientific usage of such CCDs in space.
Independently, aspects of the method (adaptation of the unbalanced Haar technique, dedicated robust linear regression) have a generic interest.
}

   \keywords{Instrumentation: detectors --
                Methods: data analysis -- 
                Techniques: image processing
               }

   \maketitle

\titlerunning{Dark signal of the SODISM frame transfer CCD in space}
\authorrunning{Hochedez et al.}

%\tableofcontents % for the HTML

%%%%%%%%%%%%%%%%%%%%%%%%%%%%%%%%%%%%%%%%%%%%%%%%%%%%%%%%%%%%%%%%%%%%%%%%%%%%%%%%%%%%%%%%%%%%%%%%%%%%%%

\section{Introduction} \label{sect:Introduction} 

When recorded for an intended astrophysical investigation or for any other scientific exploitation, raw observational data need to be corrected for various unwanted effects of instrumental nature. 
In doing so, the general goal is to reach the best possible accuracy and precision in view of a faithful estimation of the observed physical quantity. 
In the present paper, we propose a method to mitigate a couple of interrelated instrumental effects --\,the dark signal and its outliers, the hot pixels\,-- 
which affect every scientific utilization of Charge Coupled Devices (CCDs) unless they are cooled deeply. 
Our method has been designed for {frame transfer} CCDs, but some aspects of it could improve the dark signal correction of {full frame} CCDs.

The dark signal is an unwanted component of the recorded datum at each CCD picture element (pixel).
It occurs due to thermally generated electric currents in the detector and is, at first order, independent on the optical quantity of interest.

A particular motivation to improve the removal of the dark signal in CCD observational data comes from the fact that it is one of the firstly applied corrections. 
Indeed, instrumental corrections are not mutually commutative.
They must be first estimated, and then applied, in a certain logical order. 
The restoration steps proceed normally by rewinding ‘upstream’ the flow of information so that the remediation of detector effects comes before the dealing with optical issues. 
As a consequence of the early application of the dark signal correction in the sequential cleaning of the signal, a bias at this preliminary stage propagates to posterior stages of the correction process.
An under-estimation of the dark signal could \textit{e.g.} convert into an over-estimation of the scattered light.

The study that is reported in this paper is part of a larger effort that aims at enabling the scientific exploitation of the PICARD-SODISM solar images \citep{Meftah2013}. 
Therefore, the development of the method is illustrated by its application on this special set of observations.

\subsection{Charge Coupled Devices}\label{sect:CCD}

With the ever advancing CCD technology and despite the recent competition by the CMOS imaging devices, CCDs have constantly been, and still remain, detectors of choice for scientific applications 
since they were invented in 1969 at the Bell Telephone Laboratories \citep[and references therein]{Janesick1987,Smith2010}. 
Yet, their robust basic concept has remained the same: 
photocarriers --\,plus some unwanted charges generated by spurious sources\,-- are collected in a potential well created by purposely biased electrodes and/or by ion implantation that generate a fixed spatially-periodic  pattern of pixels in the semiconductor crystal. 
This is the integration phase.
At the end of it, the electric potentials of the electrodes are clocked, \textit{i.e.} varied temporally, such that the pixel pattern slides, forcing the stored electrical charges to transfer along the columns, and then along a perpendicular linear register toward an output port where the signal is converted into a voltage and somehow recorded. 
This is the readout phase.
See also Fig.~\ref{fig:CCD_scheme} for a schematic representation of the CCD concept.

The unavoidable `imperfections' of the above physical process and the corrections that are required to mitigate the resulting limitations have been studied by all manufacturers and scientific users of CCDs \citep[\textit{e.g.},][]{Rodricks2005,Burke2005}, and especially at the occasion of every space instrument embarking one \citep[\textit{e.g.},][]{Defise1997,Lo2003,Sirianni2004,Schou2004,Penquer2009,Gilard2010}.

When CCDs are included in an instrumental setup like a telescope, the first correction consists in subtracting the readout electronic offset or bias. 
This first step involves the sampling of non-physical `underscan' pixels and is relatively unmistakable. 
The assessment of the dark signal is the next necessary calibration step, and it can be a delicate one as we will see in the sections below.
There are nevertheless several other performance issues that stem from the sensing of light and its conversion to a digital image by a CCD camera. 
The list of CCD problematics includes the calibration of the {Quantum Efficiency} (QE), the knowledge of the video gain, the monitoring of non ideal flatfields, the subtraction of image residuals \citep{Rest2002,Crisp2011} and of {cosmic ray hits} (CRHs) \citep{Hill1997,Ipatov2007}, the taking into consideration of the worsening of the {Charge Transfer Inefficiency} (CTI) \citep{Rhodes2010,Baggett2012}, \textit{etc.} 

\subsection{Dark current}\label{sect:DS}

The work that is reported in this paper addresses the dark signal in frame transfer CCDs, and the dark noise that is associated to it. 
Note that the dark signal is an (unsolicited) signal, and the dark noise is the (undesired) dark signal variability, measured by \textit{e.g.} its RMS deviation (RMSD).
Furthermore, these two distinct notions must not be confused either with a third one, the {read noise} (RN), which originates in the readout port of the CCD and in the amplification stages of the analog video chain of the camera electronics. 

The dark signal that is found in a CCD pixel comes from the so-called `dark current' that is accumulated during the electronic integration and readout times. 
The dark current is due to the thermal excitation of valence electrons into the conduction band and to the simultaneous or subsequent collection of those electrons into the potential well of the pixel. 
It is said to be `dark' because it occurs even when there is no light arriving at the CCD, \textit{e.g.} when the shutter mechanism is closed. 

The dark signal is classically believed to grow linearly with the CCD integration time and to simply add arithmetically to the other sources of signal, namely the photoelectrons induced by the impinging light and the electrons produced by cosmic ray hits. 
Although some departure from linearity has been evidenced and modeled  \citep{Widenhorn2008,Widenhorn2010,Dunlap2011,Dunlap2012}, the previous linearity statements remain preponderantly valid.
It will be assumed and exploited in the proposed method.

There are mainly three sources of dark current: the `surface', the `depletion' and the `diffusion' dark current \citep{Widenhorn2002}. 
The surface dark current arises at the interface between the silicon and the oxide, below the electrodes. 
To restrain it, the {multi-pinned phase} (MPP) structure and operations are usually implemented. 
For a buried n-type channel CCD, it consists in biasing negatively the electrode phases and in adding a pinning implant under some of them.
This enables signal integration having the Si/SiO$_2$ interface fully inverted.
The holes accumulated at the interface tend to curtail the associated surface dark current by several orders of magnitude.
Yet, some surface dark current can still contribute to the signal, especially when the charges are transfered, \textit{i.e.} during the frame transfer stage if any, and during the readout stage.

The depletion --\,or `bulk'\,-- dark current is produced directly in the depletion zone, where the electrical field of the corresponding electrode determines the potential well of the pixel.
The `diffusion' dark current is generated in the field-free region below the depletion zone.
In buried channel CCDs, only the two latter components contribute significantly, but with different proportions, to the total dark signal. 

\subsection{Hot pixels}\label{sect:HP}

The dark current is not uniform. 
Quite the reverse, CCDs exhibit a number of so-called `hot pixels' that deliver spikes of dark current which can be orders of magnitude larger than elsewhere in the frame. 
Spatially, they are randomly distributed and appear as white dots in dark images.
They create an extended tail in the dark current distribution, and this tail determines chiefly the dark signal non uniformity (DSNU) \citep[\textit{e.g.},][]{Gilard2008,Gilard2010}.

A hot pixel is caused by a local discrepancy with regard to the perfect semiconducting crystal. 
It can be impurities or other crystallographic defects incorporated at the manufacturing stage.
However, the advent of hot pixels in flight must be attributed to damages caused by energetic particles, such as protons, neutrons, electrons, alpha particles, heavy ions, pions, gammas, \textit{etc.}
Those can be cosmic or solar, and many are trapped in the Van Allen belts \citep{Feynman2000}.
PICARD happens to cross the inner belt at the South Atlantic anomaly (SAA) several times per day.
This is probably the main driver for the ignition of hot pixels in its CCD.

A single event has indeed enough energy to produce either a transient ionizing effect that appears in a single frame and gets loosely labeled as a `cosmic' ray hit (CRH), whatever its origin, or permanent damage that henceforth makes the pixel hot.
In the case of inverted mode operations, granted \textit{e.g.} by an MPP architecture, the damages are believed to be mostly displacements, typically induced by non ionizing inelastic proton collisions \citep{Srour2003,Srour2006,Penquer2009}.
This is why the mechanisms by which energetic protons degrade the CTE and produce hot pixels, have been studied thoroughly for many years \citep[for example]{Hopkinson1996,Gilard2008}.
The interaction with neutrons have received attention too \citep{Chugg2003}.

Different studies have characterized the statistical behavior of hot pixels, evidencing in particular that they demonstrate the phenomenology of random telegraph signals (RTS) \citep{Hopkins1993,Hopkinson2007}.
The capacity to anneal them by `baking out' the CCD, \textit{viz.} heating it above the cold operating conditions, has also been investigated \citep[\textit{e.g.},][]{Defise1997,Sirianni2004,Polidan2004,Baggett2012}.
In parallel, p-type channel CCDs are developed with some preliminary success in an attempt to surpass the optimized radhardness of n-type channel CCDs \citep{Marshall2004,Gow2012}.
Note that, albeit a nuisance, the hot pixels can serve as a diagnostic tool to correct for the CTI \citep{Massey2010}.

\subsection{Technical solutions and data processing tools for limiting the effects of the dark current}\label{sect:DarkSolutions}

For a given CCD, after design solutions have been implemented \citep[\textit{e.g.},][]{Bogaart2009} and manufacturing care observed, the main and only true solution to limit the harmful effects of the dark signal is to cool the device. 
The dark current is indeed strongly (quasi exponentially) dependent on the temperature of the silicon crystal \citep[and Eqs.~\eqref{eq:Arrhenius} \& \eqref{eq:Widenhorn} below]{Widenhorn2002}. 
But even cooled, there is still always a residual dark signal contribution, due particularly to the fact that hot pixels are the result of impurities or crystalline defects that reveal themselves in the bulk component of the dark current, which prevails at low temperatures. 
Additionally, cooling the CCD sufficiently and regulating its temperature precisely require both important resources that are not readily available to all space instruments. 
Consequently, the CCD device cannot always be cooled as much or regulated as well as it should, even in major projects \citep{Brown2003}.
In this context, regular `bake outs' offer complementary benefits to the cold conditions of scientific acquisitions. 

Within the range of temperatures where the CCD is normally operated, the data must be corrected for the dark signal if its amplitude perturbs the measurement and if the dark signal correction is indeed susceptible to enable the intended scientific exploitation or to condition the other required instrumental corrections. 
To this aim, the typical (as well as minimal) approach is to record a dark frame just before or just after the exposed image, both taken with the same integration time, and to subtract the former from the latter. 
This method is straightforward but the mentioned subtraction increases quadratically the noise in the image of interest by the Poisson noise of the used instantaneous dark frame. 
It therefore impacts negatively the {signal to noise ratio} (SNR).
To palliate this, a master dark frame is sometimes generated from multiple dark frames. 
However, the SNR betterment offered by this master dark frame is then hampered by a lower duty cycle for the camera.
Additionally, it is often impossible to record a dark frame, and even less so the many images needed for a master dark frame, using the different integration times of all exposed images.

Various dedicated solutions have been proposed by different authors \citep[for example]{Hill1997,Widenhorn2007,Gilard2010,Cai2010}. 
Although this is not the purpose of the present work, we additionally mention the approach of \citet{Gomez-Rodriguez2009,Burger2011} who intend to not only correct for the dark signal but to also minimize the associated dark noise.

The subject of the present paper is to report about a method that has been developed to estimate and update from space data the dark signal of the frame transfer CCD onboard the PICARD-SODISM solar telescope.
Sect.~\ref{sect:Context} presents the relevant specificities of the considered payload and of its space operations. 
Sect.~\ref{sect:Estimation} presents the principle and the details of the reconstruction method which leads to the desired dark signal model (DSM).
Sect.~\ref{sect:ModelExamination} analyses and discusses the products of our DSM. 
The last section summarizes the results and concludes.

%%%%%%%%%%%%%%%%%%%%%%%%%%%%%%%%%%%%%%%%%%%%%%%%%%%%%%%%%%%%%%%%%%%%%%%%%%%%%%%%%%%%%%%%%%%%%%%%%%%%%%

\section{Description of the context} \label{sect:Context} 

%......................................................................................................................................................................................................................................................................................
\subsection{The PICARD mission and the SODISM instrument}

The imaging device which dark signal is modeled in the present paper is the flight CCD of the SODISM experiment onboard the PICARD space mission. 
PICARD was successfully launched on 15~June 2010 into a Sun synchronous dawn-dusk orbit, and commissioned in flight on 12--13~October 2010. 
It represents a European asset aiming at collecting solar observations that can serve to estimate some of the inputs to Earth climate models \citep{Thuillier2006}.
The mission scientific payload consists of the SODISM imager and of two radiometers, SOVAP (\emph{SOlar VAriability PICARD}) and PREMOS (\emph{PREcision MOnitor Sensor}),
which carry out measurements that allow estimating the {Total Solar Irradiance} (TSI) and the {Spectral Solar Irradiance} (SSI) from the middle ultraviolet to the red. 

The \emph{Solar Diameter Imager and Surface Mapper} (SODISM) \citep{Meftah2013} acquires continuously wide-field images of the photosphere and chromosphere of the Sun in five narrow pass bands centered at 215.0, 393.37, 535.7, 607.1, and 782.2\,nm.  
It contributes images that can also feed SSI reconstruction models. 
Further, the scientific objectives of SODISM encompass the probing of the interior of the Sun \textit{via} helioseismic analysis of observations in intensity on the solar disc and at the limb \citep{Corbard2008,Corbard2013}, and \textit{via} astrometric investigations at the limb.
The latter addresses especially the spectral dependence of the radial limb shape, and the temporal evolution of the solar diameter and asphericity.

Given the high metrological ambitions of the original SODISM objectives, it has been recognized that its data needed to be properly corrected against all relevant instrumental issues. 
Such corrections are indeed required to safeguard the morphometric and photometric potential of the data.
Until an optical aberration became manifest, the effects of the dark signal were expected to count among the factors that limit the most several of the scientific objectives.
This judgment triggered the present study.

The dark signal correction is specifically intended to support the scientific investigations by 
(a)~rejecting the hot pixels that disrupt the morphology of solar features, including sunspots and the radial profile of the solar limb, 
(b)~attenuating the pattern that stems from the residual dark signal non uniformity (DSNU), and particularly the striation ensuing from the memory zone component of the DSNU, 
(c)~unbiasing the subsequent estimation of the other sources of spurious signal, such as scattered light and unwanted reflections (`ghosts'), in order to minimize the photometric bias that may hamper scientific investigations. 
This last objective is quite critical because the mentioned optical effects must be estimated outside the solar disc where the various signal contaminations are all unknown, weak, and of comparable amplitudes.

In a prior stage, the SODISM telemetry is processed, together with ancillary information, to produce \emph{Level~0} (L0, or N0 for \emph{Niveau~0} in French) FITS files, including raw image data and header information.
The next stage addresses the instrumental aspects in order to generate corrected \emph{Level~1} (L1, or N1 for \emph{Niveau~1} in French) products and/or L1 correcting procedures, fed by auxiliary L1 data.
In this context, the reported work represents the first step of the general L1 endeavor, and our dark signal model is meant to supply such auxiliary L1 data.

%......................................................................................................................................................................................................................................................................................
\subsection{The SODISM CCD and its camera electronics} \label{sect:CCD_Camera} 

%......................................................................................................................................................................................................................................................................................
\subsubsection{CCD procurement and characterization by the COROT program}

The SODISM CCDs are devices from the \href{http://www.e2v.com/e2v/assets/File/documents/imaging-space-and-scientific-sensors/12-42-80.pdf}{CCD\,42-80 series of E2V}. 
They have been purchased in 2001 as part of the procurement of a batch of ten devices for the \emph{COnvection ROtation and planetary Transits} (COROT) astrometric space mission \citep{Auvergne2009}, which has been developed a few years before PICARD and was also managed by the French Space Agency (CNES). 
The main rationale for selecting a device of the 42-80 series for SODISM has been the similarity of the specifications and the project synergy with the COROT CCD program at the \emph{Laboratoire d'\'etudes spatiales et d'instrumentation en astrophysique} (LESIA) \citep{Bernardi2004,Buey2005,Lapeyrere2006,Gilard2008,Gilard2010}.

To enhance their {Quantum Efficiency} (QE) the CCD\,42-80 sensors are backthinned.
This technology improves the sensitivity, especially in the {middle ultraviolet} (MUV), {near ultraviolet} (NUV) and blue spectral ranges. 
QE and {pixel response non uniformity} (PRNU) are not further reported nor discussed here as those issues fall outside the scope of the present paper.

The selection of the four COROT flight CCDs happened on the basis of an extensive series of tests carried out by the COROT CCD program \citep{Bernardi2004,Lapeyrere2006}.

Five COROT devices that had passed the above screening became available to the SODISM  project, who additionally procured from E2V two more CCDs with identical specifications as those of the former batch (see Table~\ref{table:CCD_available_to_SODISM}).
Using the outcome of the COROT measurements and some additional characterization, the SODISM project selected for flight the CCD\,42-80-1-985 \#60 (9271-18-08), henceforth referred to as simply `CCD\#60', or SODISM `flight CCD'. 

\begin{table}[h]
\caption{List of the CCDs available to the SODISM project. CCD\#60 --\,the device selected for the SODISM flight instrument\,-- is highlighted.}
\label{table:CCD_available_to_SODISM}
\centering
\begin{tabular}{l l c}
\hline
\hline
            \noalign{\smallskip}
\textbf{Project origin}& \textbf{Reference}&\textbf{Name}\\
            \noalign{\smallskip}
\hline
            \noalign{\smallskip}
SODISM& 9274-03-06, \#78&N/A\\
            \noalign{\smallskip}
SODISM& 9274-18-06, \#89&N/A\\
            \noalign{\smallskip}
COROT& 9271-12-07, \#55&Amon\\
            \noalign{\smallskip}
\textbf{COROT}& \textbf{9271-18-08, \#60}&\textbf{Mehen}\\
            \noalign{\smallskip}
COROT& 9235-03-07, \#62&Weneg\\
            \noalign{\smallskip}
COROT& 9235-15-06, \#67&Apopis\\
            \noalign{\smallskip}
COROT& 9272-08-07, \#71&Iah\\
\hline
\end{tabular}
\end{table}

%......................................................................................................................................................................................................................................................................................
\subsubsection{Format and frame transfer architecture} \label{sect:Format}

The COROT/PICARD CCDs have a frame transfer architecture, contrarily to the regular commercial devices of the CCD\,42-80 series.
Indeed, the CCD\,42 series has the flexibility for full frame or frame transfer variants. 
The commercial devices are full frame variants only.
The reason that the full frame version is normally made rather than the frame transfer version (which could of course be operated in full frame mode) is that the additional three bus lines required up the sides of the image area to drive the image and store separately make the die size slightly larger and reduce the buttability of the array for mosaic use.
However, E2V has made the frame transfer variant for several special projects, COROT/PICARD included. 
The frame transfer architecture would allow CCD operations with no --\,or opened\,-- shutter mechanism, and/or exposing the image zone while reading out the memory zone. 
Neither of these two features are used in SODISM.
Yet, the frame transfer structure does make the dark signal issue worse than it would be for a full frame CCD as we will see in Sect.~\ref{sect:Estimation}.

The pixels of the CCD\,42-80 series are square, have a size of $13.5\times13.5\,\mu$m$^2$ and no anti-blooming structures. 
The \emph{image zone} (IZ) is $2048\times2048$.
The IZ is where the photons of interest impinge the sensor during exposure time, when the shutter is open, and where the signal accumulates during the electronic integration (see Fig.~\ref{fig:CCD_scheme}).
The \emph{memory zone} (MZ) is $2048\times2052$.
It has four extra lines as compared to the IZ because the serial register mask has four lines included with it. 
The MZ is masked with an optical aluminum shield which creates constant dark conditions there. 
This is where the signal is stored during the readout phase. 

For CCD\#60, E2V indicated that its vertical CTE was better than 0.999999, \textit{i.e} {Charge Transfer Inefficiency} (CTI = 1$-$CTE)$~\le~10^{-6}$ at {beginning of life} (BOL), and that up to $\sim$12 lines of the IZ might be covered by the storeshield of the MZ.
The LESIA tests showed that the saturation level of CCD\#60 go from 77\,$10^3$e$^-$ to 111\,$10^3$e$^-$ from the center of the image zone to its edges.
This was interpreted as an effect of the imperfect propagation of the electric potential to the electrodes at the middle of the CCD.
The CCD amplifier gain being of the order of 4\,$\mu$V/e$^-$ (See Sect.~\ref{sect:Readout} below), the CCD output voltage is in the range [0;500]\,mV.

The nominal SODISM cadence is one image per minute. 
This is also the maximal image cadence. 
Those 60 seconds include sequentially: (a) channel selection \textit{via} positioning of two filter wheels (except for dark frames), (b) CCD integration in the IZ, and associated shutter operations in case of solar exposure, (c) IZ$\rightarrow$MZ frame transfer, (d) CCD readout of the MZ, and (e) onboard processing, including image formatting and compression.

%......................................................................................................................................................................................................................................................................................
\subsubsection{\textit{A priori} estimation of the dark current at the operating temperature} \label{sect:APrioriDC}

The CCDs of the E2V\,42-80 series have a buried channel structure and they function in the {Advanced Inverted Mode Operation} (AIMO). 
AIMO is an E2V technology which brings the benefit of the {multi-phase pinned} (MPP) structure and operations --\,\textit{i.e.} negligible surface dark current\,-- while limiting the ordinary reduction in full well capacity that otherwise results from the MPP mode.

\begin{figure}[h]
%\centering
%\includegraphics[width=1.\linewidth]{SodismDark-T_CCD_evolution.png}
\resizebox{\hsize}{!}{\includegraphics{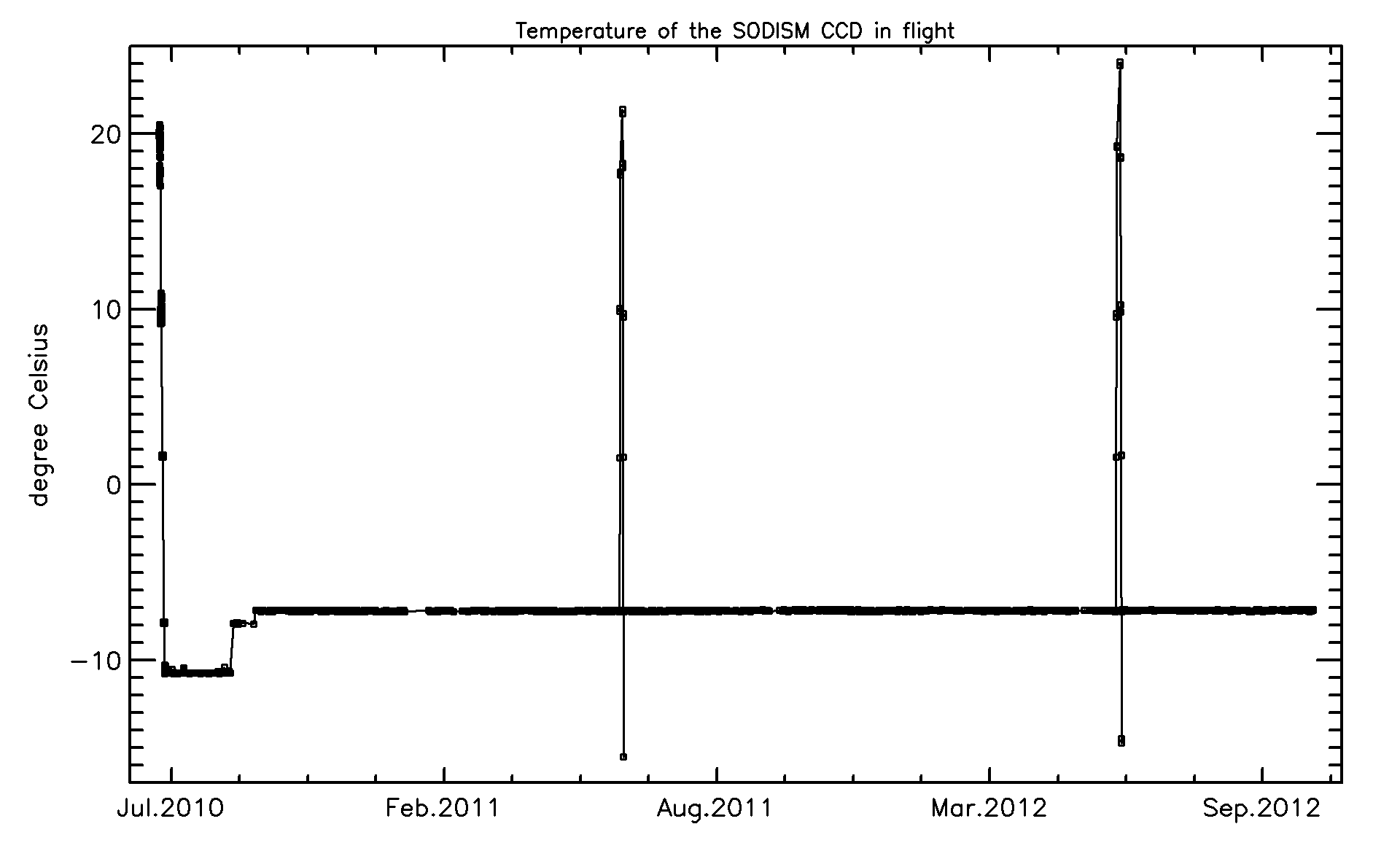}}
\caption{Evolution of the temperature of the SODISM flight CCD. After the cooldown period in July 2010, the CCD was first regulated at -11\degr C for a month and a half. 
As it was anticipated that this temperature could not be maintained for the whole duration of the mission, the setup point was subsequently raised to -7.2\degr C in mid September 2010. 
The short term and long term stabilities around that point are of the order of $\pm0.1$\degr C peak to peak (P-P). 
The precision of the measurement is 2\,mK, \textit{viz.} a hundredth of the amplitude of the residual variation. 
It can be seen that two bakeout periods occurred on 15--18 June 2011 and on 13--17 June 2012. 
They allowed heating the CCD up to 21\degr C for two days, and up to 24\degr C for three days, respectively.
Brief stages can be noticed during their warmup and cool down phases.
}
\label{fig:TCCD}
\end{figure}

The SODISM CCD is cooled so as to reach acceptable performance. 
For this, a radiator which is external to the structure of SODISM evacuates the heat from the CCD via a thermal link. 
To make the dark signal deterministic and to precisely maintain the pixel size in both the short and the long terms, the CCD is heated and regulated at $-7.2\pm0.1$\degr C peak to peak \citep[see][and Figure~\ref{fig:TCCD}]{Meftah2013}.
This is not very cold in comparison to other space experiments.
At the CCD, the regulation wipes out orbital and seasonal variations of the radiator.
Without regulation, the CCD would reach an unsteady temperature, wandering in the range {$[-16;-14]$\degr C}.
Temperature sensors are located below the Invar block supporting the CCDs. 
They provide temperature measurements with a precision of 2\,mK and an accuracy of {a few mK}.

According to \href{http://www.e2v.com/e2v/assets/File/documents/imaging-space-and-scientific-sensors/12-42-80.pdf}{the manufacturer's data-sheet}, the dark current depends on temperature as per the following Arrhenius law:
\begin{equation} \label{eq:Arrhenius}
\frac{DC(T)}{DC(293\,\mathrm{K})} = 122\,T^3\exp\left({-\frac{6400}{T}}\right) = 122\,T^3\exp\left({\frac{-E_\mathrm{g}}{2\,k_\mathrm{B}\,T}}\right)
\end{equation}
where $DC$ is the dark current at temperature $T$ [K] and $DC_0$ is the dark current at 293\,K, both expressed in $\mathrm{e}^-\cdot\mathrm{pxl}^{-1}\cdot\mathrm{s}^{-1}$ (for example). 
$E_\mathrm{g}$ is the bandgap energy of Si and amounts to $\sim$1.1\,eV; $k_\mathrm{B}$ is the Boltzmann constant ($\sim8.62\,10^{-5}$eV/K).
Note that Eq.~\eqref{eq:Arrhenius} is not in agreement with Eq.~22 of \citep{Widenhorn2002}, reproduced here:
\begin{equation} \label{eq:Widenhorn}
DC(T) = DC_\mathrm{0,diff}\,T^3\exp\left({\frac{-E_\mathrm{g}}{k_\mathrm{BT}}}\right) + DC_\mathrm{0,depl}\,T^{3/2}\exp\left({\frac{-E_\mathrm{g}}{2k_\mathrm{BT}}}\right)
\end{equation}

LESIA measured the typical dark current of \emph{non-hot} pixels of CCD\#60, and found 0.10\,$\mathrm{e}^-\cdot\mathrm{pxl}^{-1}\cdot\mathrm{s}^{-1}$ at 233K = $-40$\degr C \citep[Table~1]{Lapeyrere2006}. 
Following Eq.~\eqref{eq:Arrhenius}, the SODISM flight CCD should thus exhibit a `cool' pixel dark current ($CPDC$) around 55\,$\mathrm{e}^-\cdot\mathrm{pxl}^{-1}\cdot\mathrm{s}^{-1}\,\approx\,$4.8\,pA.cm$^{-2}$ at 293\,K $=+20$\degr C and around 4.4\,$\mathrm{e}^-\cdot\mathrm{pxl}^{-1}\cdot\mathrm{s}^{-1}$ at $-7.2$\degr C.
If we assume instead that the dark current is dominated by the second term of Eq.~\eqref{eq:Widenhorn}, non-hot pixels should exhibit a dark current of the order of $CPDC\approx3.6\,\mathrm{e}^-\cdot\mathrm{pxl}^{-1}\cdot\mathrm{s}^{-1}$ at $-7.2$\degr C, which is 80\,\% of our other estimation derived from Eq.~\eqref{eq:Arrhenius}.

In their CCD\,42-80 datasheet, E2V also specify the hottest pixels to potentially deliver up to 4\,e$^-$/pxl/hour at 153\,K. 
Following Eq.~\eqref{eq:Arrhenius}, those will deliver a dark current of about $4\,10^6\,\mathrm{e}^-\cdot\mathrm{pxl}^{-1}\cdot\mathrm{s}^{-1}$ at $+20^{o}$C, and $300\,10^3\,\mathrm{e}^-\cdot\mathrm{pxl}^{-1}\cdot\mathrm{s}^{-1}$ at $-7.2$\degr C. 
If any, the hottest pixels are therefore able to produce, during a typical one-second integration time, a dark signal that is four times the saturation level of middle CCD pixels at their nominal operating temperature of  $-7.2$\degr C.
One of our results below indicates that this apparently does not occur.

\begin{center}
\begin{table*}[t]
\centering
\caption{Parameters of the SODISM CCD camera}
\label{table:Camera}
\begin{tabular}{r l}
\hline
\hline
            \noalign{\smallskip}
\textbf{Parameter}& \textbf{Value} (at $-7.2$\degr C when relevant)\\
\hline
            \noalign{\smallskip}
CCD type&E2V CCD\,42-80 series\\
%            \noalign{\smallskip}
Flight CCD name and reference&Mehen, CCD\#60\\
%            \noalign{\smallskip}
Image zone format&$2048\times2048$\\
%            \noalign{\smallskip}
Memory zone format&$2048\times2052$\\
%            \noalign{\smallskip}
Pixel size&$13.5\times13.5\,\mu$m$^2$\\
%            \noalign{\smallskip}
CTI at BOL (beginning of life)&$\le 10^{-6}$\\
%            \noalign{\smallskip}
Saturation level&77\,--\,111\,ke$^-$\\
%            \noalign{\smallskip}
Operating temperature in flight&$-7.2\degr\pm0.1$C \\
%            \noalign{\smallskip}
Expected dark current of non hot pixel&$\sim4\,\mathrm{e}^-\cdot\mathrm{pxl}^{-1}\cdot\mathrm{s}^{-1}$ \\
%            \noalign{\smallskip}
Expected dark current of hottest pixel&300\,k$\mathrm{e}^-\cdot\mathrm{pxl}^{-1}\cdot\mathrm{s}^{-1}$ \\
%            \noalign{\smallskip}
Gain of the left readout port (image bottom)&4.22\,$\mu$V/e$^-$\\
%            \noalign{\smallskip}
Gain of the right readout port (image top)&4.14\,$\mu$V/e$^-$\\
%            \noalign{\smallskip}
Overall gain of the readout chain (16\,bit)&$\sim$\,1.18\,ADU$_\mathrm{16\,bit}$/e$^-\Rightarrow0.85\,$e$^-$/ADU$_\mathrm{16\,bit}$\\
%            \noalign{\smallskip}
Overall gain of the readout chain (15\,bit)&$\sim$\,0.59\,ADU$_\mathrm{15\,bit}$/e$^-\Rightarrow1.70\,$e$^-$/ADU$_\mathrm{15\,bit}$\\
%            \noalign{\smallskip}
Read noise (RN)&15 to 20\,e$^-$rms (increasing from BOL to EOL)\\
%            \noalign{\smallskip}
Exposure time&0.5--16\,s \\
%            \noalign{\smallskip}
Integration time&Exposure time + 0.4\,s \\
%            \noalign{\smallskip}
IZ--MZ frame transfer duration&{369.46\,ms} \\
%            \noalign{\smallskip}
Line readout duration&{11.05\,ms} \\
%            \noalign{\smallskip}
Frame readout duration&$\sim$22\,s \\
\hline
\end{tabular}
\end{table*}
\end{center}

%......................................................................................................................................................................................................................................................................................
\subsubsection{Serial register and readout ports} \label{sect:Readout}

The serial register of the E2V\,42-80 CCD is split in half. 
This allows reading out the signal at the two corresponding ports.
During nominal SODISM operations, the CCD rows are aligned with the polar axis of the Sun.
In this configuration, the North pole of the Sun is in the upper half of the image which corresponds to the `right' port of the CCD.
The lower half of the image is then readout through the `left' port.

The CCD\,42-80 output amplifier is designed to give very good noise performance at low readout rates. 
In the datasheet, E2V indicates a typical {read noise} (RN) of 3\,e$^-$rms (with a maximum of 4\,e$^-$rms) at 20\,kHz using {correlated double sampling} (CDS), and a typical amplifier gain of 4.5\,$\mu$V/e$^-$ (possibly ranging from 3 to 6\,$\mu$V/e$^-$). 

For CCD\#60, prior to delivery, E2V measured a RN of 3.9 and 4.1\,e$^-$rms for its left and right port respectively, and they indicated that the amplifier gains were equal to $3.96\,\mu$V/e$^-$ for the left port, and to $3.85\,\mu$V/e$^-$ for the right port, \textit{i.e.} a 2.9\,\% difference in favor of the left port.

In 2003, these gains were measured by LESIA in the course of the COROT screening process, and they were found to be equal to 4.18 and $4.06\,\mu$V/e$^-$ at $-40$\degr C (a 3.0\,\% difference in favor of the left port), with a temperature dependence measured to be $-820$\,ppm/K \citep[Table~1]{Bernardi2004}.
They were later reevaluated at $-40$\degr C again \citep{Lapeyrere2006}, and found to amount to 4.33 and $4.24\,\mu$V/e$^-$ respectively (\textit{i.e.} a 2.1\% difference in favor of the left port).

Following this latest result, but using the temperature dependence found by \citet{Bernardi2004}, the CCD\#60 gains are expected to be worth 4.22 and $4.14\,\mu$V/e$^-$ respectively at $-7.2$\degr C, with a discrepancy of $\sim$2\% in favor of the left port.
The exact value of the gain is also a function of the bias voltages and particularly to $V_\mathrm{OD}$, but this extra precision is not required in the sequel.

%......................................................................................................................................................................................................................................................................................
\subsubsection{Camera Electronics} \label{sect:CameraElectronics}

The first of the three main functions of the camera electronics is to supply the CCD with various bias voltages. 
% from AJV's email  of 20130220 17h02 (PIC-PR-S-7-SOD-5326-SA_Procédure de test ECA-EP_MV3.pdf, page 18):
The reset drain voltage ($V_\mathrm{RD}$) was set at {8.42}\,V.
The substrate voltage ($V_\mathrm{SS}$) was set at {0.0}\,V. 
The output gate voltages ($V_\mathrm{OG1}$ \& $V_\mathrm{OG2}$) were set at {$-$6.60}\,V and {$-$5.51}\,V, respectively. 
The output drain ($V_\mathrm{OD}$) and the dump drain ($V_\mathrm{DD}$) voltages were set at 21.78\,V and 14.18\,V, respectively. 

Secondly, the camera must sequence the CCD and the shutter mechanism with the needed clocks. 
It should be noted that the exposure defined by the mechanical shutter occurs strictly \emph{within} the duration of the electronic integration, which always exceeds the former by 0.4\,s for this reason. 
Therefore, the integration of dark signal lasts longer than the integration of photo-electrons from the solar exposure by 400\,ms. 
The shutter exposure duration can be programmed to take any value between 0.5\,s (minimum) and 16.0\,s (maximum), by increments of 0.1\,s, and the integration time goes correspondingly from 0.9\,s to 16.4\,s. 
Both durations are given in SODISM FITS headers.

When the integration is over, the signal that has accumulated in the \emph{image zone} (IZ) is transfered to the \emph{memory zone} (MZ) in {369.46\,ms}. 
This is the frame transfer and it occurs shutter closed, \textit{viz.} under dark conditions.
Note that at this point, there can be CTE issues occurring, as well as some evolving contribution from surface dark current which might add a column dependent pedestal.

Thirdly, the camera electronics amplifies and converts the 0\,--\,500\,mV analog signals that must be read out from the two output ports of the CCD (See Sect.~\ref{sect:Format}) into a stream of digital numbers. 
The conversion cadence being 100\,kHz pixel, the CCD line rate is {11.05\,ms} per row, and the total readout duration amounts to 22\,s for a whole frame that is read from the two CCD ports simultaneously.
The digital stream is then directed to the {Picard Gestion Charge Utile} (PGCU) computing unit, where the image is formed, processed, and most often, compressed. 

The gain of the preamplifier stage was fixed to $9\,$V/V.
This is attenuated by the impedance adaptation by a factor 0.94, which leads to an effective gain of $8.46\,$V/V.
The unipolar input range of the {analog-to-digital converter} (ADC) is 0\,--\,2.8\,V and it operates on 16\,bit.
This implies a conversion factor of $2^{16}/2.8$ {analog-to-digital unit} (ADU) per Volt, \textit{i.e.} $2.34\,10^{-2}$\,ADU/$\mu$V. 
Hence, the overall gain of the readout chain is:
\begin{equation}\label{eq:G16}
\begin{split}
G_{16} & \approx2.34\,10^{-2}\mathrm{ADU}/\mu\mathrm{V}\times8.46\times4.2\,\mu\mathrm{V/e}^- \\
& \approx0.83\,\mathrm{ADU}_\mathrm{16\,bit}/\mathrm{e}^-,\\
\mathrm{\textit{i.e.}}~G_{16}^{-1} & \approx\,1.20\,\mathrm{e}^-/\mathrm{ADU}_\mathrm{16\,bit},
\end{split}
\end{equation}
with a discrepancy of about 2\% between the two CCD ports.
However, most SODISM data are coded on 15\,bits, the sixteenth and {least significant bit} (LSB) being simply dropped.
The gain is then halved and becomes:
\begin{equation}\label{eq:G15}
\begin{split}
G_{15} & \approx0.42\,\mathrm{ADU}_\mathrm{15\,bit}/\mathrm{e}^-, \\
\mathrm{\textit{i.e.}}~G^{-1}_{15} & \approx2.40\,\mathrm{e}^-/\mathrm{ADU}_\mathrm{15\,bit}.
%$\sim$0.45\,ADU/e$^{-}$, \textit{i.e.} $\sim$2.2\,e$^{-}$/ADU in the instrument paper
\end{split}
\end{equation}

To verify the above bottom-up estimation of the gain, we estimate it now by means of the `photon transfer technique' \citep{Janesick1987a,Downing2006} or, more precisely, by using a `dark signal transfer technique'.
It states that the random variables of the dark signal, of its standard deviation, and of the read noise are related by:
\begin{equation}\label{eq:DarkSignalTransfer}
\left( \sigma_S\mathrm{[ADU]} \right)^2=G\mathrm{[ADU/e}^-]\times S\mathrm{[ADU]}+\left( RN\mathrm{[ADU]} \right)^2
\end{equation}
Obviously, this is valid if the noise sources are solely shot noise and read noise.
It does not apply to the hot pixels which exhibit erratic fluctuations and intermittency.
When $RN$ becomes negligible with respect to the dark Poisson noise, Eq.~\eqref{eq:DarkSignalTransfer} turns into:
\begin{equation}\label{eq:DarkSignalTransfer2}
\begin{split}
G^{-1}\mathrm{[e}^-\mathrm{/ADU]}& \approx S/\sigma_S^2 \mathrm{~or,}\\
\log{G^{-1}}& \approx\log{S}-2\times\log{\sigma_\mathrm{S}}
\end{split}
\end{equation}

\begin{figure}[h]
%\begin{figure*}[t]
%\sidecaption
%\includegraphics[width=12 cm,clip=true,trim=0mm 0mm 200mm 0mm]{SodismDark-PhotonTransfer.png}
\resizebox{\hsize}{!}{\includegraphics[clip=true,trim=0mm 0mm 220mm 0mm]{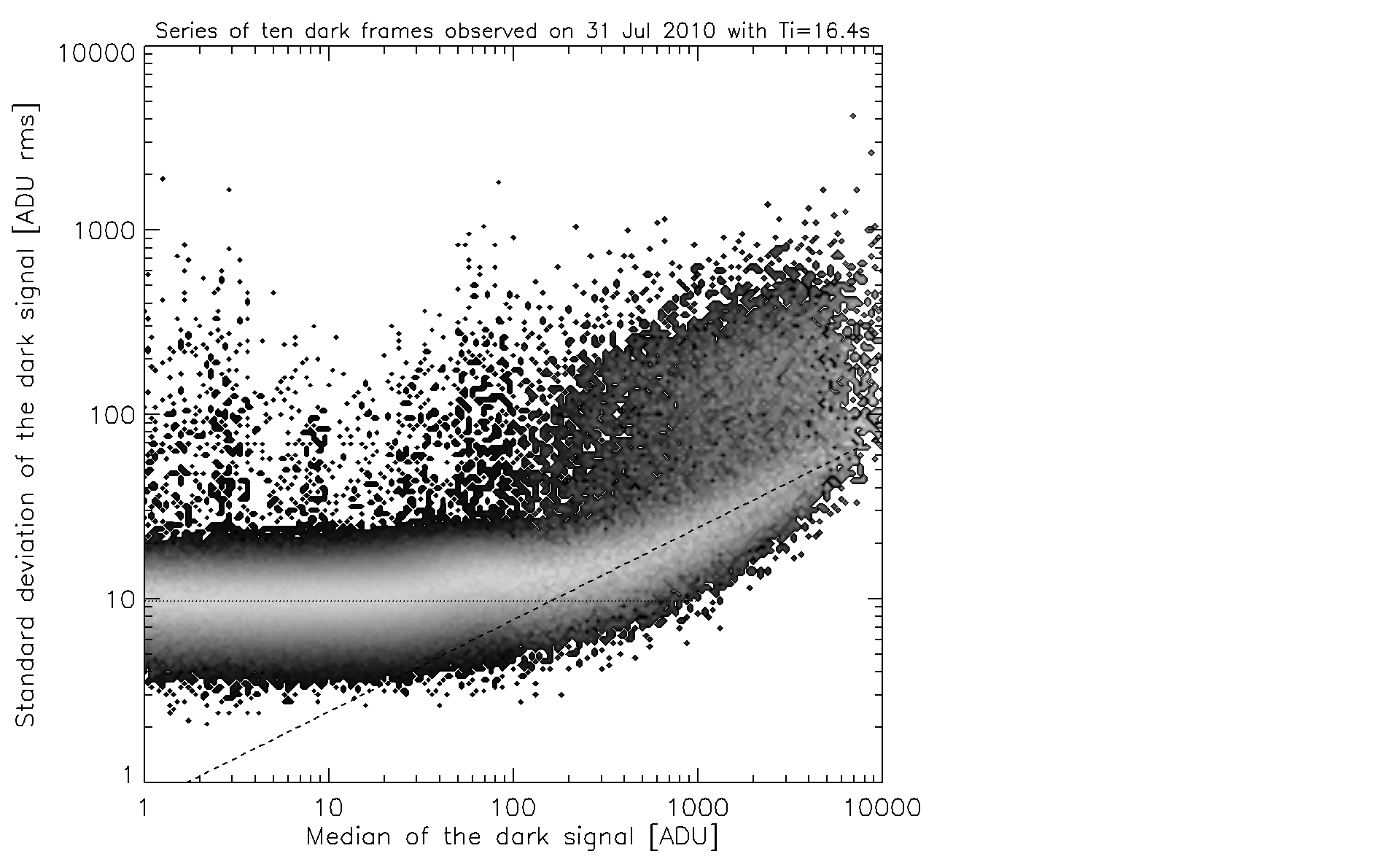}}
\caption{Dark signal transfer plot. The median and standard deviation are computed for every pixel of a series of ten dark frames acquired on 31 July 2010 and represented here as a 2-dimensional histogram. 
A region on the left is dominated by the read noise (RN), which amount to about 10\,ADU$_\mathrm{15\,bit}$rms. 
The photon shot noise dominated region has a slope of 0.5 in the logarithmic representation.}
\label{fig:PTC1}
\end{figure}

On 31 July 2010, ten dark full frames were acquired on 15\,bits, with a 16.4\,s integration time, and within few hours. 
For each pixel, after having subtracted the offsets, we compute the median and the standard deviation and plot their two-dimensional histogram (Figure~\ref{fig:PTC1}).
The read noise dominated region and the shot noise dominated region are well distinct.

%\end{figure*}

\begin{figure}[h]
\resizebox{\hsize}{!}{\includegraphics{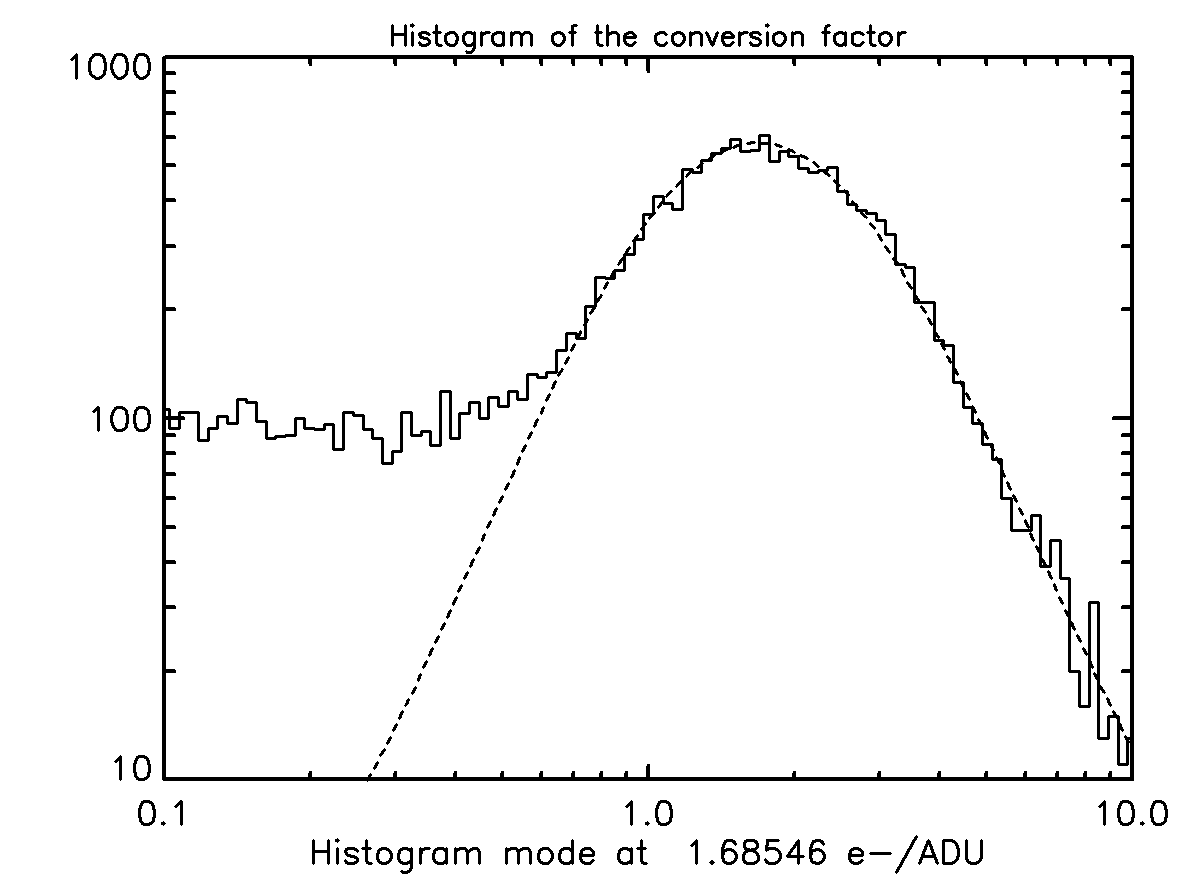}}
\caption{Histogram of G$^{-1}_\mathrm{15\,bit}$, the inverse of the video gain. Each pixel in the shot noise dominated region of Figure~\ref{fig:PTC1} provides an estimate of G$^{-1}_\mathrm{15\,bit}$. 
A Gaussian function is fitted and gives 1.685$\pm$0.1\,e$^-$/ADU$_\mathrm{15\,bit}$ for the mode.}
\label{fig:PTC2}
\end{figure}

According to Eq.~\eqref{eq:DarkSignalTransfer2}, $\log{G_{15}^{-1}}$ can therefore be estimated by measuring the mode of the histogram of $[\log{(Median)}-2\log{\sigma}]$ (Figure~\ref{fig:PTC2}).
Fitting a Gaussian function to this distribution gives a value of 1.685$\pm$0.1\,e$^-$/ADU$_\mathrm{15\,bit}$ for $G^{-1}_{15}$.
This is hardly consistent with the bottom-up estimation of 2.40\,e$^-$/ADU$_\mathrm{15\,bit}$  in Eq.~\eqref{eq:G15}.
The discrepancy may be explained by the approximate knowledge regarding \textit{e.g.} the impedances used in the bottom-up approach.
We will henceforth use the values of the gain measured by our in-flight analysis, which also provides an estimate for the {readout noise} (RN), worth 9.7\,ADU$_\mathrm{15\,bit}$, that is $\sim$16\,e$^-$rms, in July 2010, at BOL.
This is about four times more than the 4\,e$^-$rms value expected from the CCD alone (Sect.~\ref{sect:Readout}).

\begin{figure}[h]
\resizebox{\hsize}{!}{\includegraphics[clip=true,trim=0mm 0mm 0mm 20mm]{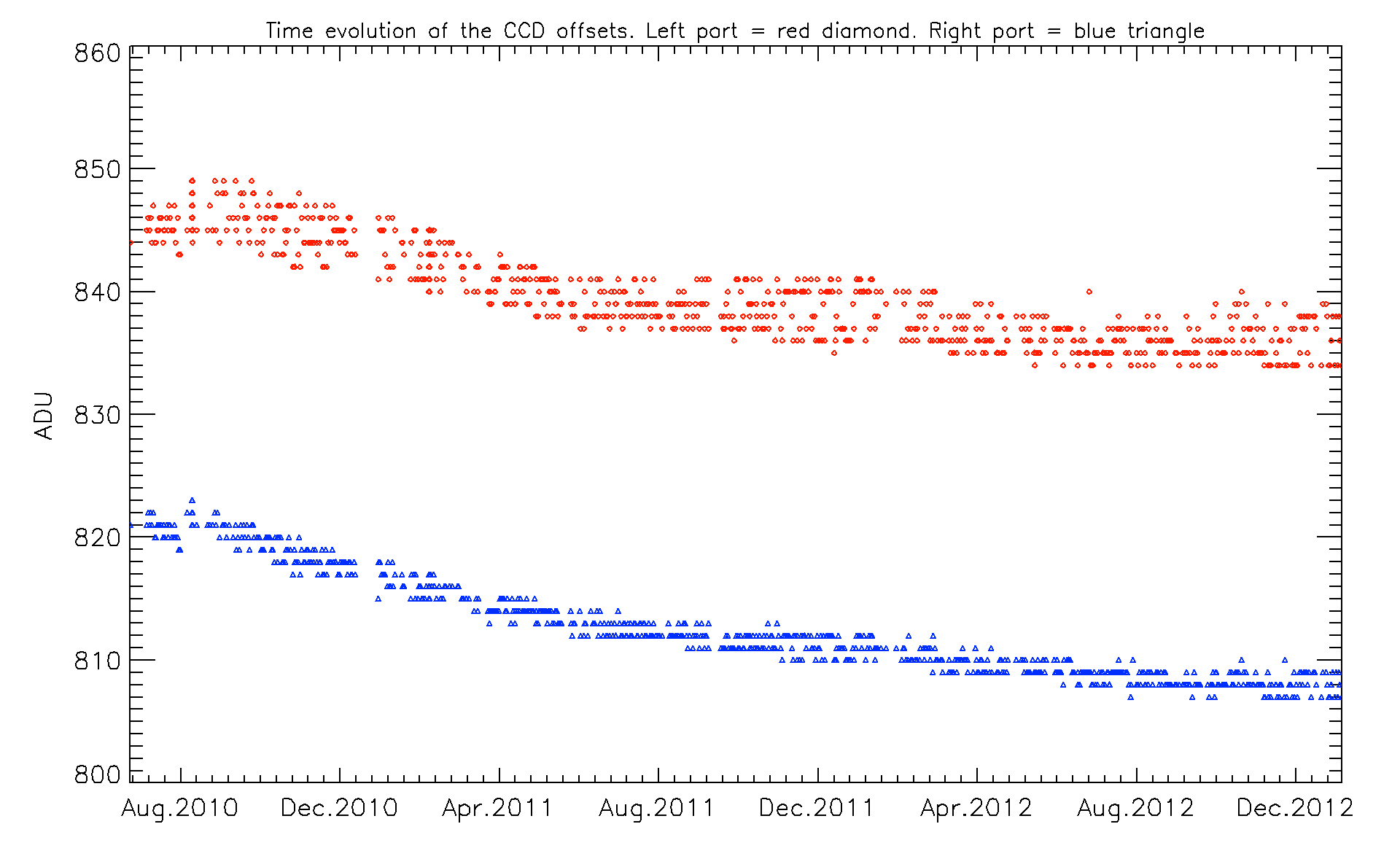}}
\caption{Temporal evolution of the offsets of the left port (red diamonds) and right port (blue triangles) of SODISM CCD from July 2010 to December 2012. 
The offset value is larger and the spread is larger for the upper time series which corresponds to the left port.}
\label{fig:Offsets} 
\end{figure}

In order to avoid negative values that the ADC would not be able to convert, a voltage bias, or \emph{offset}, is added to the signal before it is fed to the ADC.
Its value is in the [840--850]\,ADU$_\mathrm{15\,bit}$ range for the `left' port of the CCD, and in the [810--820]\,ADU$_\mathrm{15\,bit}$ range for its `right' port (Figure~\ref{fig:Offsets}). 
These two offset values are estimated onboard for each image by calculating  the average of an underscan area.
The (rounded) values are later transmitted in the {telemetry} (TM) and appended to the FITS header of the L0 image products. 
It is noticed that the time series of the offsets exhibit some spread and it was demonstrated (not shown in the present paper) that this variability is dominated by an orbital modulation.
This coupling may be linked with the observed RN excess noted above.

As the standard deviations of the same underscan area are also computed, this allows estimating RN. 
It increases slowly from $\sim9\,\mathrm{ADU}_\mathrm{15\,bit}\mathrm{rms}\,\approx15\,\mathrm{e}^-$\,rms at BOL, up to $\sim11\,\mathrm{ADU}_\mathrm{15\,bit}\mathrm{rms}\,\approx20\,\mathrm{e}^-$\,rms at EOL.

The main results of Sect.~\ref{sect:CCD_Camera} are summarized in Table~\ref{table:Camera}.

%......................................................................................................................................................................................................................................................................................
\subsection{Available in-flight dark frames}

As part of the automatic ordinary science operations, also know as the `SODISM routine' \citep{Meftah2013}, it had been foreseen before launch to record one dark frame per day with 1.4\,s integration time.
This value is close to the nominal integration time of most channels at BOL.
But it has been recognized during the first weeks of flight operations that dark frames with other integration times needed to be acquired.
Indeed, for the 535\,nm channel that is dedicated to helioseismology (535H), the exposure time is equal to 7\,s.
Additionally, at the occasion of special campaigns, the exposures in any channel can be different than 1\,s.
Finally, the expected degradation of the UV channels was to lead to successive increases of their exposure time, which, in 2012, has reached 16\,s (the maximum value allowed by the camera electronics) for the 215\,nm channel, and 6\,s for the 393\,nm channel.

\begin{figure}[h]
\resizebox{\hsize}{!}{\includegraphics{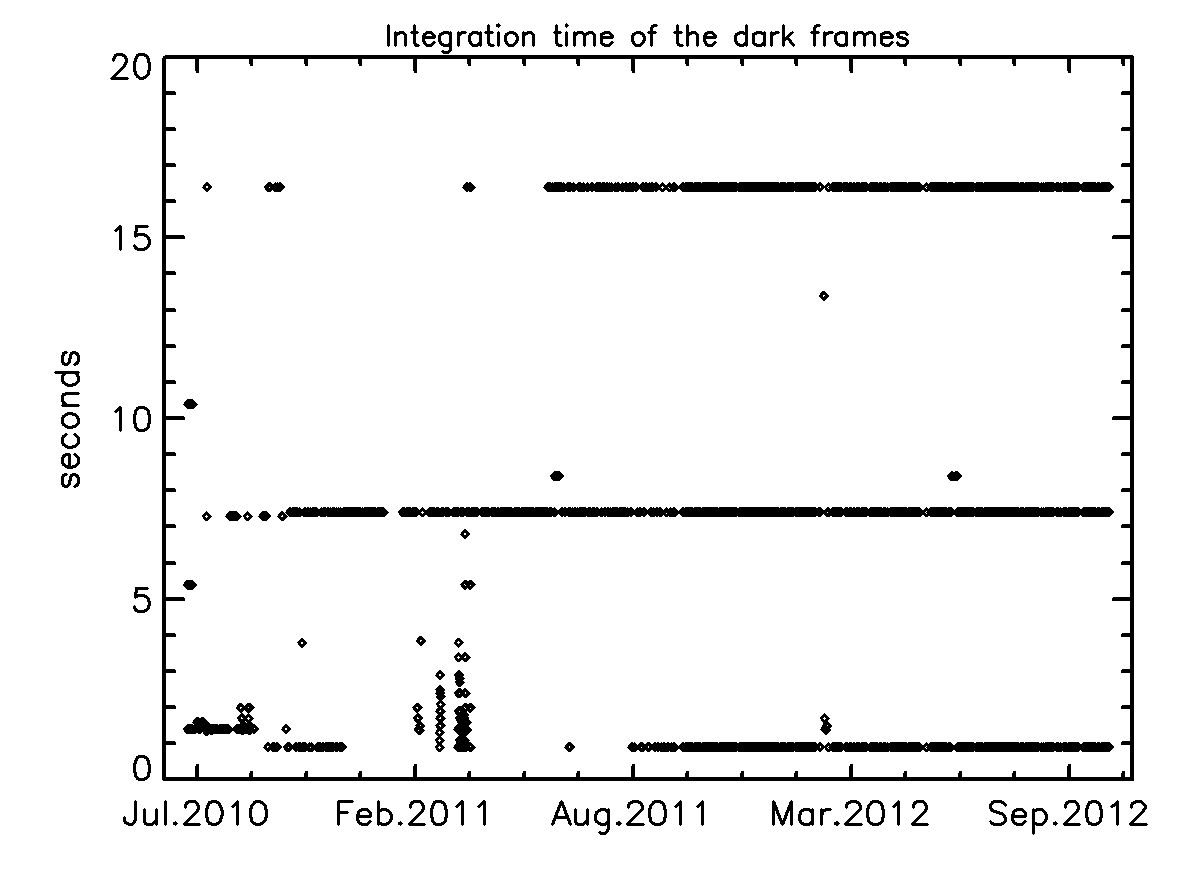}}
\caption{Evolution of the integration time of the full CCD dark frames that are available to generate the desired dark signal model (DSM). 
It can be noticed that the dark frame integration time pattern was not regular during the first year of the mission, 
until end of August 2011, after which three dark frames are recorded each day, with integration time equal to 0.9, 7.4 and 16.4\,s. 
Before the summer of 2011, most days and many weeks exhibit no alternation of the dark frame integration time.}
\label{fig:IntegrationTime} 
\end{figure}

However, PICARD is a micro-mission and some time has been required to modify the coded routine. 
This is why the integration time of the daily dark frames had to be alternated manually (between 0.9\,s and 7.4\,s) in autumn 2010 (25 Sept.\,--\,2 Dec. 2010). 
Later, in February and March 2011, special linearity campaigns have permitted to vary the dark integration time at few occasions.
But for the rest, the dark exposure time was set to 7.4\,s until 8 June 2011 when it could be again alternated manually among two values (7.4, and 16.4\,s) for three months, and then automatically from 25 August 2011 onwards, among three values (0.9, 7.4, and 16.4\,s). 
The dark frame program has thus been pretty irregular during the first year of the mission when other aspects of the performance were at their best (Figure~\ref{fig:IntegrationTime}).

It should be mentioned that other dark signal images have been acquired $\sim$\,50 times per day from 5 Aug. 2010 until 27 Mar. 2012. 
However, they offer information only within a ring shape spanning across the solar limb, and until now, those were not exploited by the present analysis because we are interested in estimating a model of the dark signal over the whole CCD frame.

%%%%%%%%%%%%%%%%%%%%%%%%%%%%%%%%%%%%%%%%%%%%%%%%%%%%%%%%%%%%%%%%%%%%%%%%%%%%%%%%%%%%%%%%%%%%%%%%%%%%%%

\section{Modeling the dark signal from in-flight data} \label{sect:Estimation} 

Once modeled, the dark signal correction consists simply in a subtraction since we will assume linear behavior with respect to:
\begin{itemize}
      \item its addition to the (wanted and spurious) components of the signal, and
      \item integration time proportionality.
\end {itemize}

But as a prerequisite, before performing this subtraction, we need to model, \textit{viz.} to be able to estimate, the CCD dark signal for any exposure duration and at any instant of time during the mission lifespan.
This section describes a method that delivers such a dark signal model (DSM) on the basis of in-flight dark observations scheduled with only few different integration times.

%......................................................................................................................................................................................................................................................................................
\subsection{Image zone and memory zone dark signal components}

\begin{figure}[h]
\resizebox{\hsize}{!}{\includegraphics{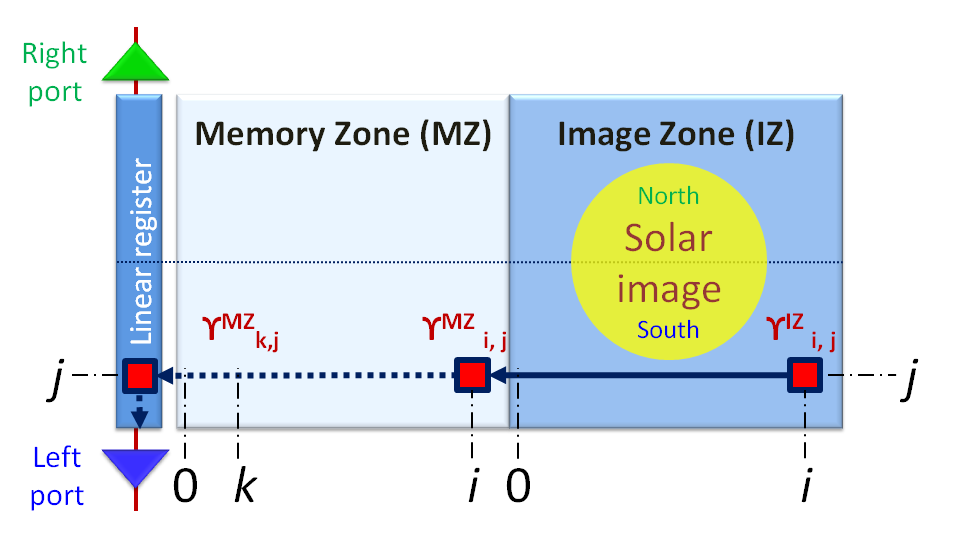}}
\caption{Sketch of a frame transfer CCD and scenario for the dark signal build up. 
A frame transfer CCD, such as SODISM's, has two zones, IZ and MZ having the (quasi-) same format. 
During the CCD integration time, the signal is generated and stored in the Image Zone (IZ) on the right of the figure. 
This is when the shutter can be opened and the device exposed to observe the Sun. 
This is also when the IZ component of the dark signal accumulates at the rate of $\Upsilon^{IZ}_{i,j}(t)\,\mathrm{e}^-\cdot\mathrm{pxl}^{-1}\cdot\mathrm{s}^{-1}$.
The image frame is then transfered to the Memory Zone (MZ). 
At this occasion, the signal charges that were generated at coordinates ($i,j$) of the IZ are moved to ($i,j$) of the MZ. 
During the next and last phase, the readout phase, the CCD lines (represented vertically here) are transfered one by one to the serial register. 
The readout phase lasts 22\,s, which is sufficient to accumulate a significant amount of extra dark signal while sojourning in the $i$ pixels ($k,j$) of the MZ.}
\label{fig:CCD_scheme}
\end{figure}

Regarding frame transfer CCDs, the main consideration pertains to the generation of dark current in both the {image zone} (IZ) and the {memory zone} (MZ).
The dark signal is indeed the addition of an IZ component that is proportional to the programmable integration time, with an MZ component that is the sum of the contributions of all physical pixels of the MZ where the quantity of interest sojourns during readout (see Fig.~\ref{fig:CCD_scheme}).
Practically, a pixel of coordinates $(i,j)$, having thus accumulated signal charges in row~$i$ and column~$j$ of the IZ, transfers rapidly its signal, just before the start of readout, to the pixel of same coordinates in the MZ.
This `IZ--MZ' frame transfer takes only {$\sim$370}\,ms and is expected to add little extra dark signal.
Once in MZ, the quantity of interest gets shifted $i$~times toward the serial register.
During this slow process, it accumulates supplementary dark signal in each of the $i$~physical pixels of the MZ where it dwells for {11.05\,ms} while one of the foregoing CCD lines is read out.

The above scenario is formalized in Eq.~\eqref{eq:DSds} and \eqref{eq:DarkSignalLong}:
\begin{align} 
DS_{i,j}(t) &= O_{i,j}(t) + ds_{i,j}(t), \label{eq:DSds}\\
ds_{i,j}(t) &= G\cdot\left[ (T+\delta T)\cdot\Upsilon^\mathrm{IZ}_{i,j}(t) + \tau\cdot\sum_{k=0}^{i}{\Upsilon^\mathrm{MZ}_{k,j}(t)}\right], \label{eq:DarkSignalLong}
\end{align} 
where,
\begin{itemize}
\item $DS_{i,j}(t)$ is the random variable of the dark measurement recorded at date $t$ for the pixel of coordinates $(i,j)$. Note that the available dark images have always been recorded with 15\,bit resolution.
\item $O_{i,j}(t)$ is the random variable of the offset in this pixel at this time. Apart from the CCD port dependence, we will henceforth consider that it does not depend on the pixel as we do not have access to this information. 
Hence $O_{i,j}(t)=O(t)$ and we use the amount that is computed onboard and provided in the L0 FITS header for the given CCD port (see Sect.~\ref{sect:CameraElectronics}).
\item $ds_{i,j}(t)$ is the random variable of the dark signal at $t$ for the pixel $(i,j)$. 
\item $G$ is the gain of the video chain (see Sect.~\ref{sect:CameraElectronics} and Eqs.~\eqref{eq:G16} and \eqref{eq:G15}).
\item $T$ is the programmed duration of the shutter exposure, also known as the \emph{exposure time}.
\item $T+\delta T=T'$ is the duration of the CCD integration, also known as the \emph{integration time}. As mentioned in Sect.~\ref{sect:CameraElectronics}, $\delta T=0.4\,$s.
\item $\Upsilon^{IZ}$ is the random variable of the dark current map in the \textit{image zone} (IZ) at date $t$.
\item $\tau$ is the duration needed to read a CCD line. $\tau=11.05$\,ms.
\item $\Upsilon^{MZ}$ is the random variable of the dark current map in the {memory zone} (MZ) at date $t$.
\end{itemize}

Eq.~\eqref{eq:DarkSignalLong} can then be rewritten into:
\begin{equation} \label{eq:DarkSignalShort}
ds_{i,j}(t) = G \cdot \left[ T' \cdot \Upsilon^{IZ}_{i,j}(t) + \tau\cdot \Psi^{MZ}_{i,j}(t)\right]
\end{equation}
with, 
\begin{equation} \label{eq:Ramp}
\Psi^{MZ}_{i,j}(t)=\sum_{k=0}^{i}{\Upsilon^{MZ}_{k,j}(t)}
\end{equation}

For every pixel, there are two unknowns, $\Upsilon^{IZ}$ and $\Psi^{MZ}$. 
In other words, we need to reconstruct the dark signal contributions of both the IZ and the MZ.
This is a linear regression problem and it can be solved if there are at least two different integration times $T'$ for a given state of $\Upsilon^{IZ}$ and $\Psi^{MZ}$.

Before detailing the process of reconstruction, we notice in Eq.~\eqref{eq:Ramp} that $\Psi^{MZ}_{i,j}$ is a sum of positive contributions and must consequently be growing with $i$.

Let us also recall that a pixel is cool (\textit{viz.} non-hot) until it gets hit by an ionizing particle and turns hot (see Sect.~\ref{sect:HP}). 
The $ds_{i,j}(t)$ time series are thus expected to display temporally piecewise constant behaviors. 
They are not necessarily always growing with time because hot pixels may cool down even if they appear to never come back cool.

%......................................................................................................................................................................................................................................................................................
\subsection{Generation of the dark signal model}

\subsubsection{Global concept}

The goal is to estimate at any time $t$, $\Upsilon^{IZ}_{i,j}(t)$ and $\Psi^{MZ}_{i,j}(t)$, the cartographies of the IZ and MZ dark signal components.
Given that Eq~\eqref{eq:DarkSignalShort} depends only on unknowns that are associated to a single pixel $(i,j)$, we proceed by processing the temporal dependence column per column to spare computer memory, and then, pixel per pixel.

First, since integration time T'\,=\,7.4\,s is the configuration that has been programmed most regularly (see Fig.~\ref{fig:IntegrationTime}), a provisional dark signal model (DSM) will be established for this particular value of T'.
The `DSM@7.4s' model will result from the fitting of piecewise constant functions to the observed time series because the ignition and cool down of hot pixels --\,whether in IZ or MZ\,-- are sudden events.

Secondly, for every pixel, we will identify the time intervals during which the dark signal of this pixel --\,and so, both of its IZ and MZ components\,-- are constant according to DSM@7.4s.

Thirdly, for each such interval and for each pixel, we perform a robust regression \textit{versus} integration time.
Next, this enables updating the values of $\Upsilon^{IZ}_{i,j}(t)$ and $\Psi^{MZ}_{i,j}(t)$ for the successive periods of constancy.

Finally, the $\Upsilon^{IZ}_{i,j}(t)$ and $\Psi^{MZ}_{i,j}(t)$  data cubes are resampled on a daily grid by means of a median filter.

%\paragraph{Generation of a dark signal model at fixed integration time.}
The next three sections (Sect.~\ref{section:BoxCox7.4s} to Sect.~\ref{section:UHT7.4s}) describe the formation of `DSM@7.4s'.
Its outcome is then employed in Sect.~\ref{section:Regress_All}.

\subsubsection{Step 1 -- Variance stabilization of the time series at T'=7.4\,s}\label{section:BoxCox7.4s}
The unbalanced Haar transform (UHT) that is going to be needed in Step 3 (Sect.~\ref{section:UHT7.4s}) requires the time series to be made approximatively homoscedastic.
This is why the $ds^{7.4\,s}_{i,j}(t)$ of Eq.~\eqref{eq:DSds} (with $t$ chosen so that $T'(t)=7.4$\,s) are first of all preprocessed by a generalized Box-Cox power transform \citep{Box1964,Sakia1992}:
\begin{equation} \label{eq:Box-Cox}
ds^\mathrm{7.4\,s}_{\mathrm{BoxCox}~i,j}(t)=\frac{\left( ds^{7.4\,s}_{i,j}(t)+\alpha \right)^\lambda-1}{\lambda \cdot \mathrm{GM}(ds^{7.4\,s}_{i,j})^{\lambda-1}},
\end{equation} 
where GM is the geometric mean.

Indeed, it can easily be derived from Eq.~\eqref{eq:DarkSignalTransfer} that --\,for the cool pixels\,-- $ds^\mathrm{7.4\,s}_{\mathrm{BoxCox}~i,j}(t)$ will be homoscedastic if:
\begin{equation}\label{eq:BoxCox}
\begin{split}
\lambda & =0.5 \mathrm{~and,}\\
\alpha & =RN^2\times G^{-1}\approx 10^2\times1.70=170\,\mathrm{ADU}_\mathrm{15\,bit},
\end{split}
\end{equation}
which reminds of \citet{Anscombe1948} too. 
For the Jul. 2010 -- Nov. 2012 period, the smallest $ds^{7.4\,s}_{i,j}(t)$ observation has been $-178$\,ADU$_\mathrm{15\,bit}$, which appears in agreement with the 
%Eq.~\eqref{eq:BoxCox} 
$\alpha$ value that is required to variance-stabilize the quadratic addition of read noise and shot noise.
The above determined Box-Cox transform will of course be less effective on hot pixels, but it \textit{a posteriori} appears to be sufficient for the purpose of the UHT.

\begin{figure}[h]
%\sidecaption
\begin{center}
(a) After Step 1
\includegraphics[width=\linewidth]{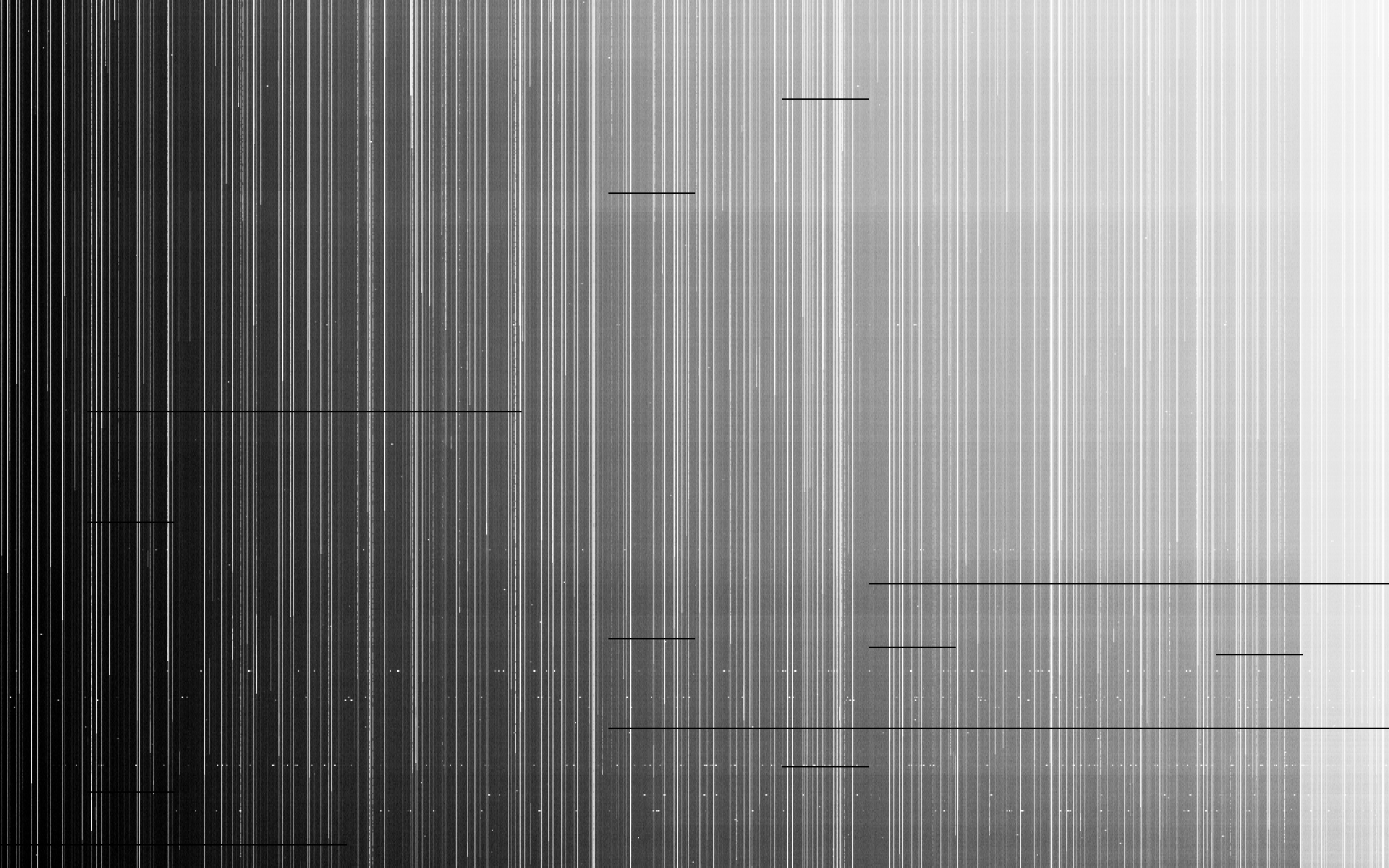}
(b) After Step 3 
\includegraphics[width=\linewidth]{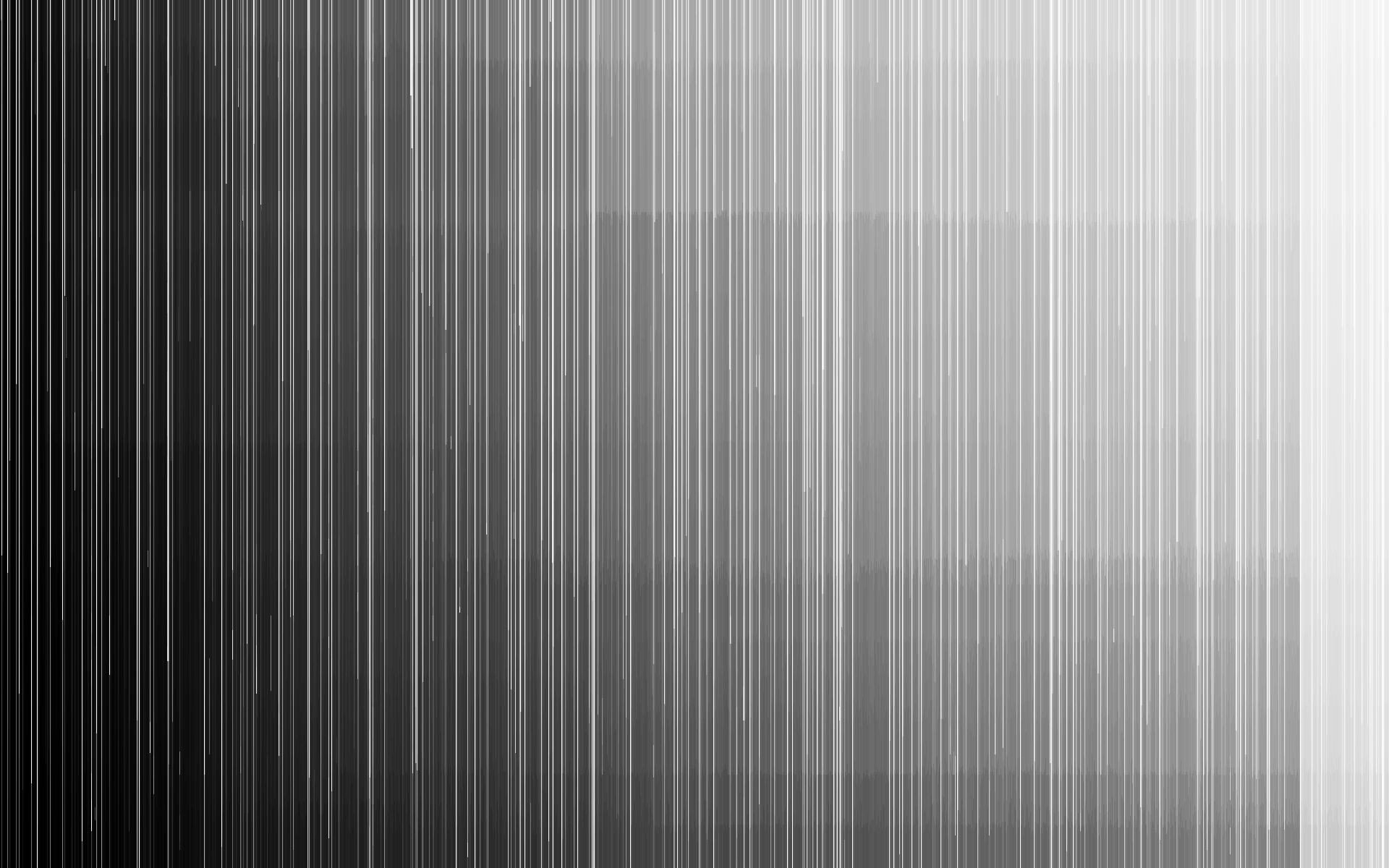}
\end{center}
\caption{Upper figure (a): evolution of the dark signal in CCD column (image row) \#1341 for an integration time T' equal to 7.4\,s. 
     The Box-Cox transformed dark signal of this CCD column are represented horizontally (after Step 1), and the information from all frames having T'=7.4\,s has been stacked vertically.
     The presence or advent of hot pixels can be seen as bright semi-infinite vertical lines, telemetry holes as black horizontal segments, cosmic ray hits as white dots.
     It can be verified that the pixels on the right part of the figure exhibit higher dark current since they dwell for a longer duration in the memory zone.
     We notice that in this particular column \#1341, a hot pixel in the MZ generates a dark signal excess affecting the last $\sim$10\% of the CCD column, on the right of the figure.
     Lower figure (b): the same data, but after Step~3.
}
\label{fig:AfterStep1and3} 
\end{figure}

The Box-Cox preprocessed time series of CCD column \#1341 is represented in Fig.~\ref{fig:AfterStep1and3}-a.

\subsubsection{Step 2 -- Median filtering of the time series at T'=7.4\,s}\label{section:MedianFilter7.4s}

Real observational data are cluttered with various outliers. 
In particular, telemetry missing blocks (TMBs) and cosmic ray hits (CRHs) introduce spiky features in the time series of most pixels.
Both of those are fortunately easy to correct for by using a median filter that flags the signal when it departs from the running median by more than 5 running $\sigma$.
When this occurs, the information is declared erroneous and replaced by the running median.

\begin{figure}[h]
\includegraphics[width=\linewidth]{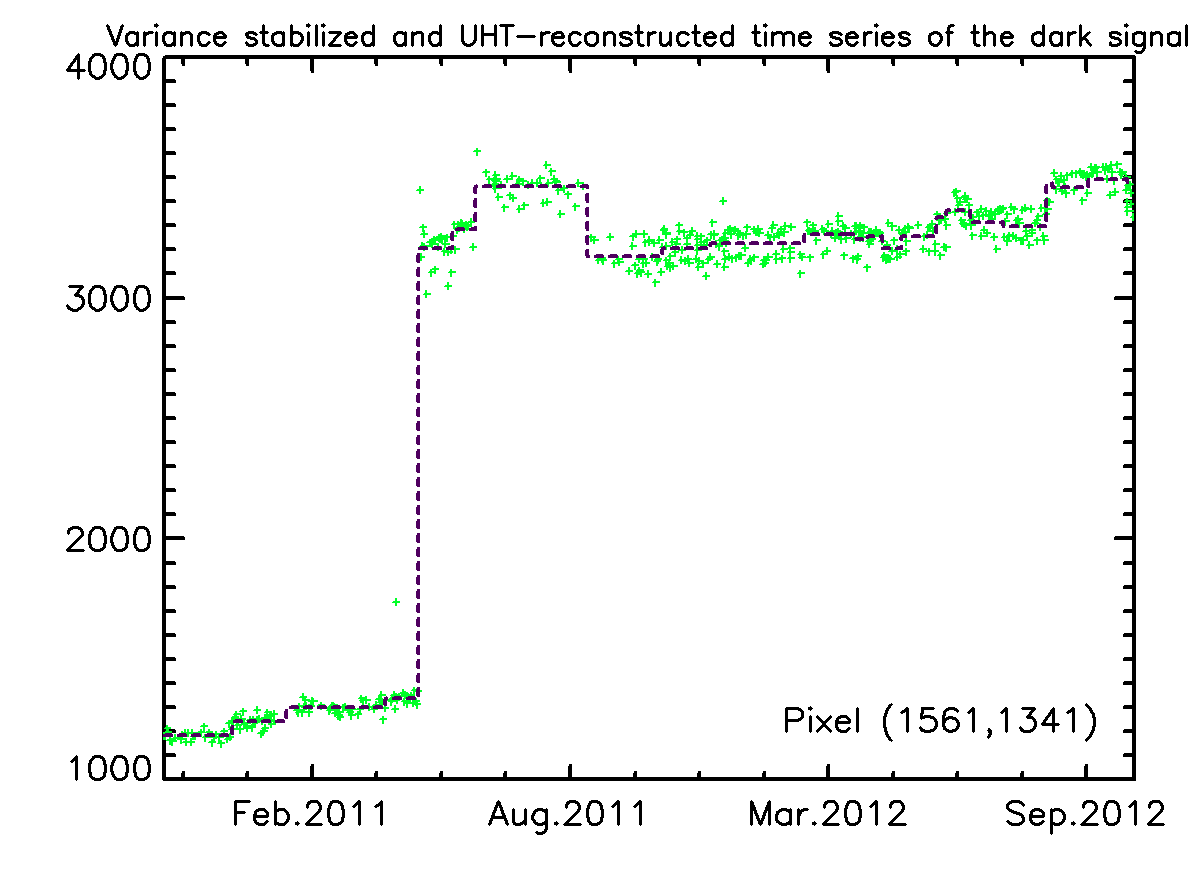}
\caption{Temporal evolution of the variance stabilized dark signal (T'=7.4\,s) in pixel (1561,1341), over-plotted with the piecewise constant fit (dashed line) that results from the reconstruction by the unbalanced Haar technique. 
This represents a vertical cut in Fig.~\ref{fig:AfterStep1and3}-a and the corresponding one in Fig.~\ref{fig:AfterStep1and3}-b.
Note that the dark signal appears to oscillate between two values after the pixel has turned hot.
Such multi-level random telegraph noise (RTS) is expected \citep{Chugg2003,Hopkinson2007}.
}
\label{fig:Pixel1561_1341EvolutionUHT} 
\end{figure}

\subsubsection{Step 3 -- Unbalanced Haar transform of the time series at T'=7.4\,s} \label{section:UHT7.4s}

The sought dark signal models, and `DSM@7.4s' in particular, are anticipated to be piecewise constant (\textit{viz.} staircase) functions of time.
Indeed, the advent of hot pixels or their partial cooldown, either in image zone (IZ) or memory zone (MZ), should create perfect steps in the DSM@7.4s time series.
If in the IZ, the transition will be of relatively large amplitude, while the ignition of a hot pixel in the MZ is expected to generate a much smaller jump (with a ratio of $\tau$/T' $\approx$~0.14\%).
This discrepancy will however be attenuated by Step~1 and it remains logical to look for an algorithm that fits staircase functions with no \textit{a~priori} on the instant of the steps, their amplitudes, nor on their number.
Reciprocally, no change in the dark signal should occur outside interactions with an energetic particle and annealing events. 

There are several such algorithms and we selected the multiscale `unbalanced Haar transform' (UHT) described by \citet{Fryzlewicz2007} because ``the jumps in the basis functions do not necessarily occur in the middle of their support, [...] which avoids the restriction of jumps occuring at dyadic locations'' \citep{Fryzlewicz2007}, and because its computational complexity is of order $O(n \log{n})$. 
Also, \citet{Fryzlewicz2007} demonstrates the outstanding performance of his UHT scheme \citep[see][Figure~1, for example]{Fryzlewicz2007}.

Variance stabilization is a prerequisite to the UHT technique, and this motivated Steps 1 and 2.
By decomposing on a Haar-like basis the median-filtered time series that come out from Step 2, the UHT provides the desired instants of time in-between which no hot pixel transition has taken place for the considered pixel.

Moreover, the UHT enables denoising and reconstructing a piecewise constant time series that fits the input.
Fig.~\ref{fig:AfterStep1and3}-b represents the temporal evolution of CCD column \#1341 after Step 3. It can be compared with Fig.~\ref{fig:AfterStep1and3}-a.

For a selected pixel that ignited in the Spring of 2011, Fig.~\ref{fig:Pixel1561_1341EvolutionUHT} shows the temporal evolution of its variance-stabilized dark signal.
The UHT staircase function is overplotted.
The prevalent goodness of the UHT fit can be appreciated in this particular case.

We have reprogrammed the UHT for the Interactive Data Language (IDL) on the basis of the R package that \citet{Fryzlewicz2007} makes \href{http://cran.r-project.org/web/packages/unbalhaar/index.html}{available online}.
A thresholding of the unbalanced Haar-like wavelet coefficients is needed to not reconstruct the input completely, but to preserve only the significant steps, and to hence denoise.
We define it as per Eq.~\eqref{eq:UHT_threshold}, which has been devised empirically to single out the physical signal.
\begin{equation} \label{eq:UHT_threshold}
\lvert w \rvert \times scale^{2.25} > 4\,10^4,
\end{equation}
where $w$ is the coefficient of the considered UH wavelet, and $scale$ is the length of the shortest of its two piecewise constant segments.
Clearly, this forces short steps to display jumps of large amplitudes, while this permits longer periods of stability to result from smaller leaps.

\subsubsection{Step 4 -- Robust regression for constant dark signal periods}\label{section:Regress_All}

Having identified the instants of time when a given IZ pixel ignites, or when it changes level if it was already hot, or when any pixel of its corresponding MZ column discloses a significant jump, 
it becomes possible to discriminate $\Psi^{MZ}$ and $\Upsilon^{IZ}$, the MZ and IZ dark components respectively, by means of a linear regression on the integration time.
This is the purpose of Step 4 where we consider all dark frames that have been acquired with different integration times $T'$.

The pixel coordinates $(i,j)$ being fixed, and $t$ being in a time interval $\Delta$ when $\Psi^{MZ}_{i,j}(t)$ and $\Upsilon^{IZ}_{i,j}(t)$ can be supposed constant, Eq.~\eqref{eq:DarkSignalShort} suggests that a linear regression \citep[see \textit{e.g.}][as a possible reference]{Kutner2004} will allow separating the two.
The regression must however be robust against the heteroscedasticity of the data that cannot be variance-stabilized anymore since we want to exploit their proportionality to T'.
To this aim, we use a scheme that employs a weighted L$_1$ norm, as defined in Eq.\eqref{eq:L1} below.

We define $ds_\Delta(t)=ds(t)$ for which $t \in \Delta$. 
Further, for each of the $K$ distinct $T'_k$, we gather the dark signals $ds^k_\Delta(t)$ defined by the $ds_\Delta(t)$ for which $T'(t)=T'_k$, and we compute:
\begin{itemize}
\item their median: $MED^k_\Delta=\,\mathrm{median}_{t}(ds^k_\Delta(t))$,
\item their mean absolute deviation with respect to the median: $MAD^k_\Delta=\,\mathrm{median}_{t}\left( \lvert ds^k_\Delta(t) - MED^k_\Delta \rvert \right)$, and
\item the maximum of all standard deviations estimated from Eq.~\eqref{eq:DarkSignalTransfer}: $MSD^k_\Delta=\max_t(G\cdot ds^k_\Delta(t)+ RN^2)^{0.5}$.
\end{itemize}

As it is known that $\sigma\approx 1.4826 \times \mathrm{MeanAbsDev}$ \citep{rousseeuw2005}, we define $\sigma^k_\Delta=\max[MSD^k_\Delta,1.4826 \cdot MAD^k_\Delta]$ in order to tentatively improve the standard deviation estimate for the hotter pixels.
A robust linear regression (RLR) is then obtained by solving iteratively:
\begin{align} 
(\Upsilon_\Delta^\mathrm{RLR},\Psi_\Delta^\mathrm{RLR})_{i,j} & =\underset{\Upsilon>0,\Psi>0}{\arg\min}\,\mathrm{D}_1 \left( ds_\Delta,[\Upsilon,\Psi] \right) \label{eq:argmin} \\
\mathrm{with,}\nonumber\\
\mathrm{D}_1 \left( ds_\Delta,[\Upsilon,\Psi] \right) & = \frac{1}{K}\,\sum_{k=1}^K \left( \frac{\lvert ds^k_\Delta-(\Upsilon\times T'_k+\Psi)\rvert}{\sigma^k_\Delta} \right) \label{eq:L1}
\end{align} 

In case $K=1$, the RLR is still fed with the available $ds^k_\Delta(t)$, but also with $\Psi_{Previous\,\Delta}^\mathrm{DSM}$ (the last DSM-computed value for $\Psi$), associated with a null integration time (T'=0).

\subsubsection{Step 5 -- Updating and resampling on a daily grid}\label{section:UHT_All}

\begin{figure}[h]
\begin{center}
(a) CCD column \#1341 Image Zone (IZ) DSM evolution
\includegraphics[width=\linewidth]{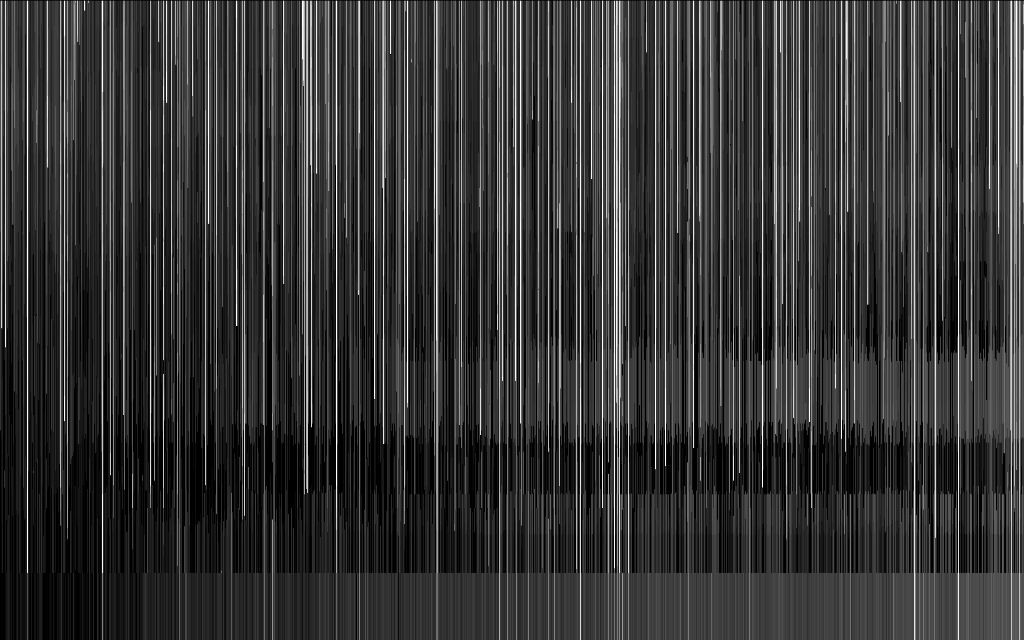}
(b) Memory Zone (MZ) DSM evolution for the same column
\includegraphics[width=\linewidth]{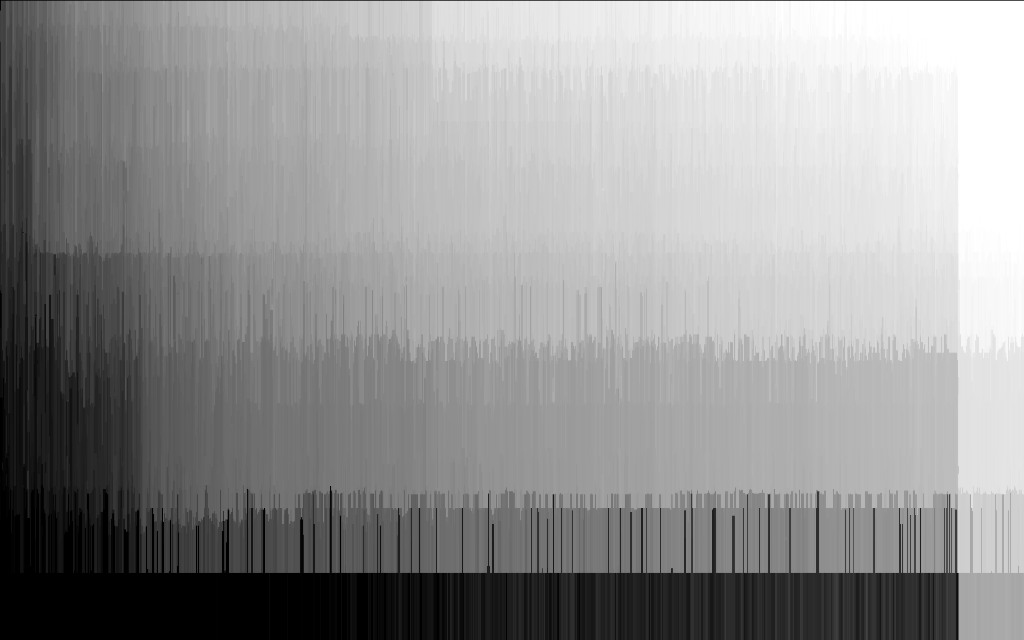}
\end{center}
\caption{Upper figure (a): evolution of the IZ dark signal model (DSM) for CCD column (image row) \#1341. 
     The log of the IZ DSM is represented horizontally, and the information from all frames has been stacked vertically.
     The advent of hot pixels in the IZ appears as bright semi-infinite vertical lines.
     Lower figure (b): evolution of the Memory Zone (MZ) DSM for the same column. 
     The representation conventions are the same as above, but contrarily to figure (a), the gray scale is linear.
     The presence or advent of hot pixels in the MZ is seen as steps that brighten the figure rightward and upward simultaneously.
     The dark current in the MZ appears as an inclined plane, which slope increases with time due to the gradual emergence of a distribution of hot pixels in the MZ.
     For both IZ and MZ, the first weeks (bottom of the images) are not well estimated because there were no data to feed the DSM@7.4 for those dates.
}
\label{fig:AfterStep5} 
\end{figure}

We finally need to acknowledge that the intervals $\Delta$ are dissimilar in their capacity to inform about $(\Upsilon_\Delta,\Psi_\Delta)_{i,j}$.
For this reason, the $(\Upsilon_\Delta^\mathrm{RLR},\Psi_\Delta^\mathrm{RLR})$ that has been computed in Step 4 cannot simply be assigned to the period of time covered by $\Delta$.
The following updating scheme is used to adequately mix a fresh estimate computed in $\Delta$ by Step 4, with the previous one, already assigned to $Previous\,\Delta$.

A quality index $Q_{\Delta,(i,j)}$ is computed for each $\Delta$:
%\begin{align}
\begin{equation}
\begin{array}{ll}
Q_1 &=\max_k T'_k - \min_k T'_k \\
Q_2 &= \sqrt{K} \\
Q_3 &= \exp{\left( -\lvert \, \log \mathrm{D}_1 ( ds^\Delta,[\Upsilon^\Delta,\Psi^\Delta] ) \,\rvert \right)} \\
Q^\Delta &= Q_1 \times Q_2 \times Q_3\\
\end{array}
\end{equation} 
$Q_1$ and $Q_2$ naturally favor the leverage provided by unalike and/or numerous $T'_k$ respectively.
$Q_3$ measures both the goodness and the reliability of the Step 4 robust linear regression (RLR).
Note that $Q_3$ reaches a maximum for D$_1 ( ds^\Delta,[\Upsilon^\Delta,\Psi^\Delta] ) =1$.
This is the desirable behavior since the terms of the sum in Eq.~\eqref{eq:L1} must obviously not be much larger than 1, if the fit is to be good.
Yet, they must not be much smaller than 1 either, as this hints at poor statistics due to a lack of data.
In case $K=1,~Q_\Delta$ takes its previous value, arbitrarily divided by 50.

Now, $\Upsilon$ and $\Psi$ can be updated by taking the weighted mean of their new and old estimations, with their respective quality index as coefficients.
The quality index is also updated, but using the geometric mean.
\begin{equation}
\begin{array}{l}
\medskip
\Upsilon_\Delta^\mathrm{DSM}= \left({Q_\Delta \cdot \Upsilon_\Delta^\mathrm{RLR}+Q_{Previous\,\Delta}^\mathrm{DSM} \cdot \Upsilon_{Previous\,\Delta}^\mathrm{DSM}}\right) /
  \left({ Q_\Delta+Q_{Previous\,\Delta}^\mathrm{DSM}}\right) \\
\smallskip
\Psi_\Delta^\mathrm{DSM}= \left({Q_\Delta \cdot \Psi_\Delta^\mathrm{RLR}+Q_{Previous\,\Delta}^\mathrm{DSM} \cdot \Psi_{Previous\,\Delta}^\mathrm{DSM}}\right) /
  \left({ Q_\Delta+Q_{Previous\,\Delta}^\mathrm{DSM}}\right) \\
Q_\Delta^\mathrm{DSM}= \sqrt {Q_{Previous\,\Delta}^\mathrm{DSM} \cdot Q_\Delta} \\
\end{array}
\end{equation} 

\begin{figure}[!h]
\begin{center}
(a)~Image zone dark current, 3 Nov. 2012
\includegraphics[width=.9\linewidth]{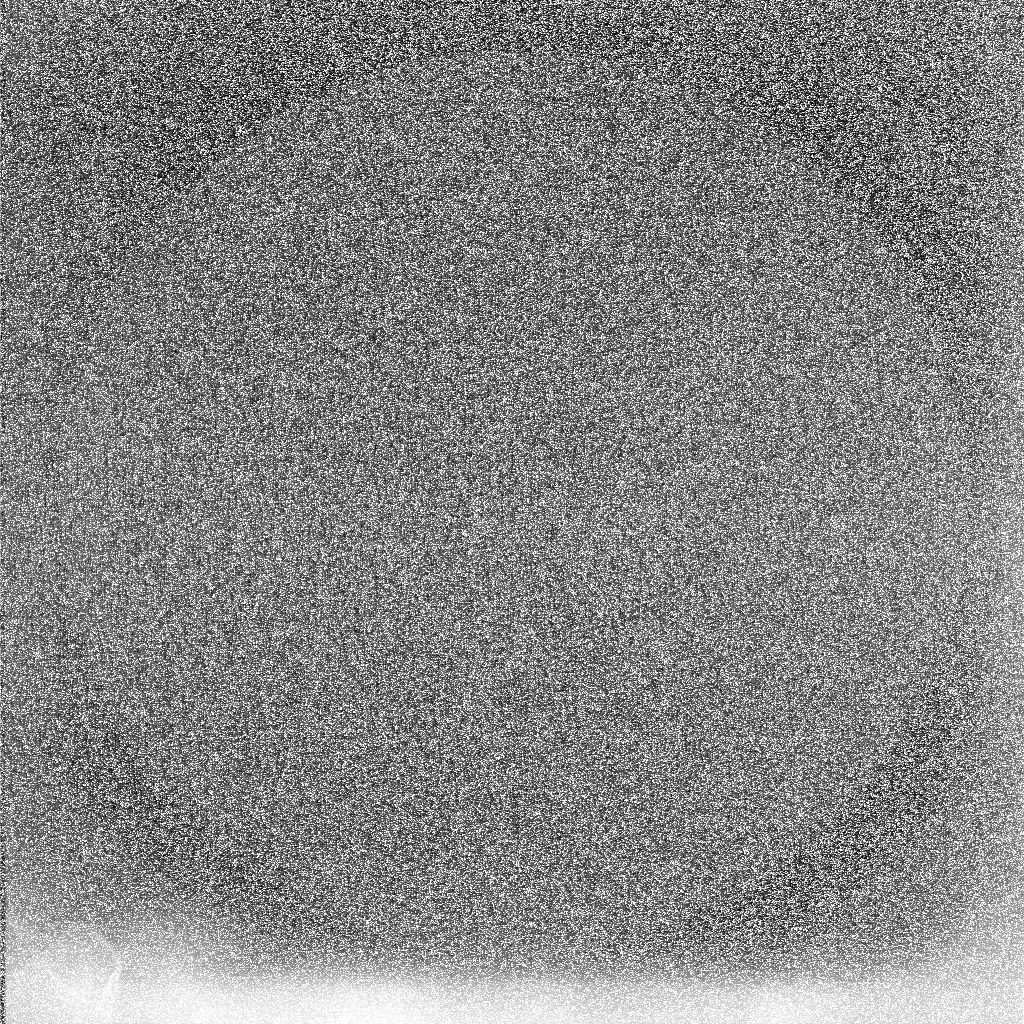}
(b)~Memory zone dark signal, 3 Nov. 2012
\includegraphics[width=.9\linewidth]{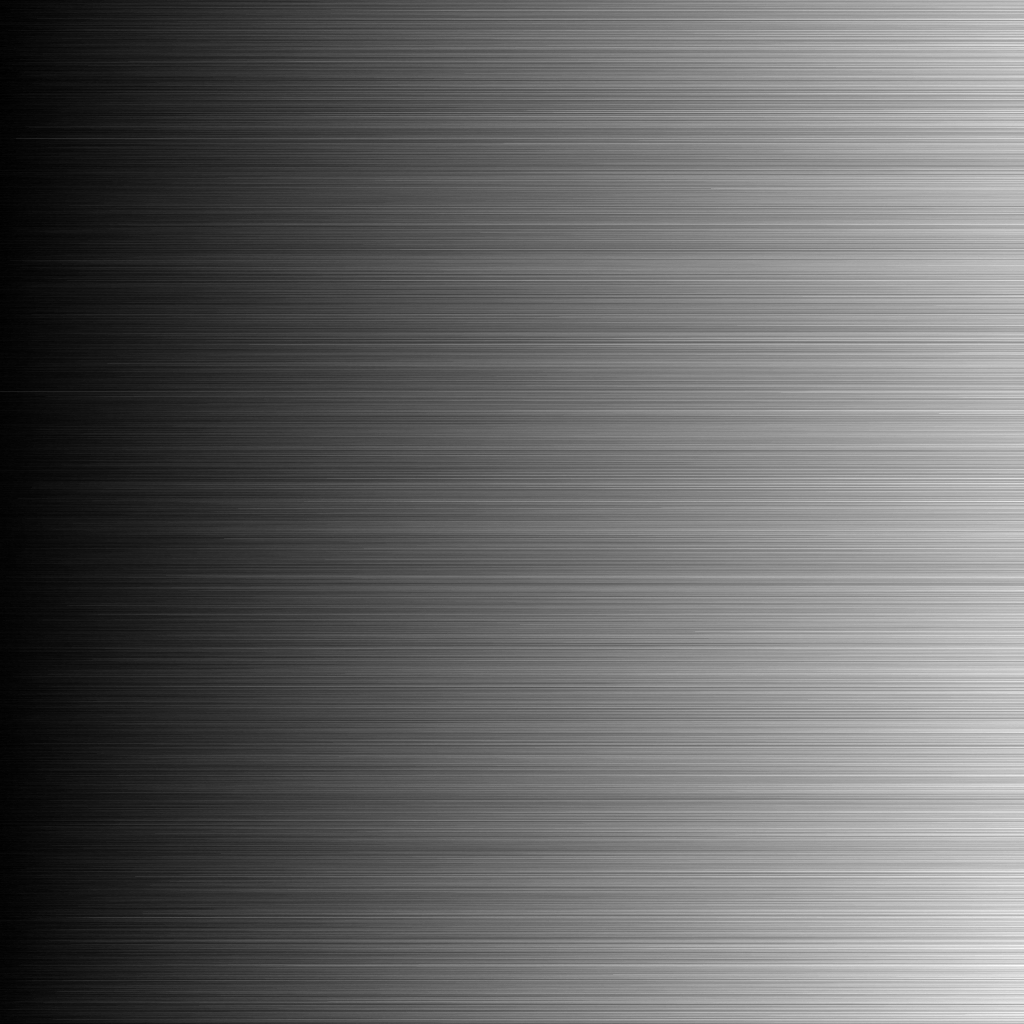}
\end{center}
\caption{
Upper figure (a): DSM dark current in the image zone as of 3 Nov. 2012. 
The gray scale is logarithmic. Black corresponds to 2\,ADU/s and white to 30\,ADU/s.
At the bottom of the image, one can notice an extended contamination by a spurious signal.
Its level is in the 6--20\,ADU$\cdot\mathrm{pxl}^{-1}\!\cdot\mathrm{s}^{-1}$ range (see also Fig.~\ref{fig:ZI_Histogram}).
A persistent solar image is also apparent in the IZ dark current.
Its amplitude is about 0.2\,ADU$\cdot\mathrm{pxl}^{-1}\!\cdot\mathrm{s}^{-1}$.
Lower figure (b): dark signal in the memory zone reconstructed by the DSM as of 3 Nov. 2012. 
The gray scale is linear and displays the quasi linear increase of the dark signal in MZ (from left to right in this representation).
Black corresponds to 20\,ADU and white to 700\,ADU.
}
\label{fig:ZI_ZM} 
\end{figure}

Finally, in order to provide a straightforward dark signal correction scheme, daily IZ and MZ matrices are computed pixel per pixel by taking the daily median of the updating stage outputs if any, or by duplicating the values of the day before otherwise.
Fig.~\ref{fig:AfterStep5} shows the outcome of Step 5 for CCD column \#1341, the same column as in Fig.~\ref{fig:AfterStep1and3}.
Fig.~\ref{fig:ZI_ZM} shows the outcome of Step 5 for the whole IZ and MZ as of 3 Nov. 2012.
Note that at this stage, a binary mask of the hot pixels is additionally constructed by thresholding the IZ cartographies with a value (\textit{viz.} 50\,e$^-\!\cdot\mathrm{pxl}^{-1}\!\cdot\mathrm{s}^{-1}$) that is legitimated in Sect.~\ref{sect:IZ_distribution}.

The resulting daily FITS files are used by the L1 processing chains of the Picard Science Mission Center (CMS-P), and distributed 
\href{http://picard.projet.latmos.ipsl.fr/data.php?dir=CO.models\%2FN1B\%2F}{online} from the SODISM homepage at LATMOS, which is located at \url{http://picard.projet.latmos.ipsl.fr/}.

%%%%%%%%%%%%%%%%%%%%%%%%%%%%%%%%%%%%%%%%%%%%%%%%%%%%%%%%%%%%%%%%%%%%%%%%%%%%%%%%%%%%%%%%%%%%%%%%%%%%%%

\section{Model examination} \label{sect:ModelExamination} 

In the present section, the products of the above described model are investigated. 
We measure in Sect.~\ref{sect:Validation} the discrepancy between our dark signal model (DSM) and a number of observed dark frames, and we verify that the bias is small.
Secondly, we study in Sect.~\ref{sect:IZ_distribution} the temporal evolution of the estimated IZ dark current. 
A few additional verifications are reported at the end of this section.

\subsection{Checking the dark signal model (DSM) on dark frames} \label{sect:Validation} 

\begin{figure}[h]
\begin{center}
(a) Evolution of the histograms of 7.4\,s corrected dark frames
\includegraphics[width=\linewidth]{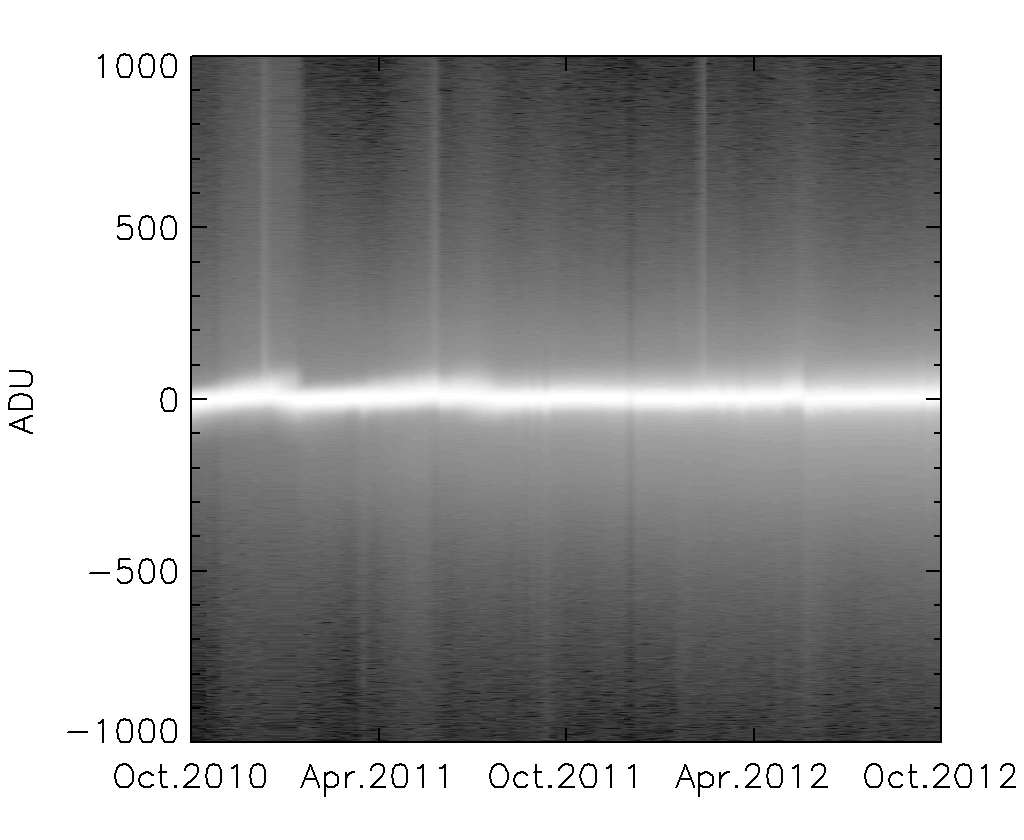}
(b) Average histogram and Gaussian fit
\includegraphics[width=\linewidth]{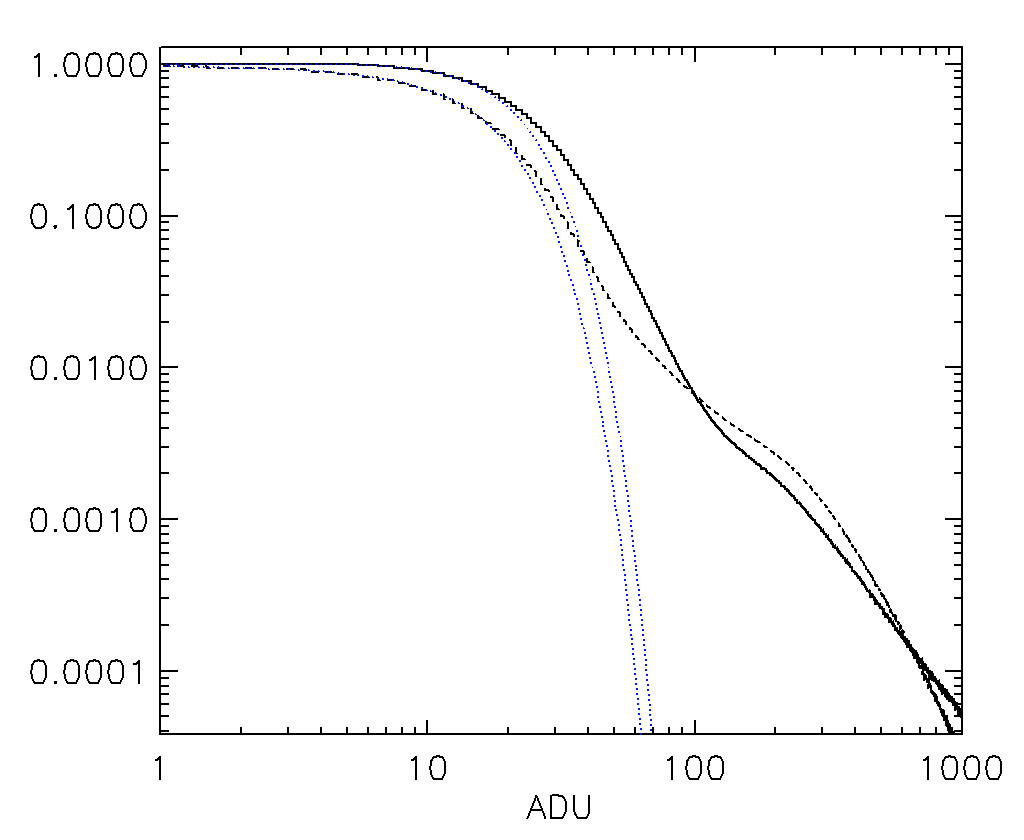}
\end{center}
\caption{
Upper figure (a): Temporal evolution of corrected dark frame histograms. 
Dark frames having weekly regularity and $T'$=7.4\,s integration time have been selected.
Their histogram is coded with a logarithmic gray scale.
Most pixels display a corrected signal near 0, as expected.
Lower figure (b): Average histogram of all above dark frames.
Both axis have log scales. The number of occurrences corresponding to negative signals are represented by the dash line.
The fitted Gaussian function is overplotted with a blue dotted line.
Its center (the model bias), its standard deviation (model error), and its full width at half maximum (FWHM) are worth $\sim$3.1\,ADU $\sim$14.7\,ADU, and $\sim$34.6\,ADU respectively.
}
\label{fig:DarkCorrectionEvolution} 
\end{figure}

The goal is here to inspect the outcome of the model in the situation for which the solution that the DSM should ideally deliver is known.
We select dark frames programmed with the 7.4\,s integration time, \textit{i.e.} the most common configuration.
It is the integration time of the images recorded in the 535\,H channel that is dedicated to the helioseismologic investigations which are demanding in term of photometric correction.
This also helps getting a regular sampling, set weekly hereafter.

All along the mission, the corrected dark signal appears to be tightly distributed around zero, as it should (Fig.~\ref{fig:DarkCorrectionEvolution}-a).
The few lighter vertical strips are related to frames having many cosmic ray hits (CRHs) taken \textit{e.g.} while crossing the South Atlantic anomaly (SAA), and obviously, the DSM does not model the CRHs.
By fitting a Gaussian function to the histogram (Fig.~\ref{fig:DarkCorrectionEvolution}-b), we learn that the DSM underestimates the dark signal by $\sim$3.1\,ADU $\approx$5\,e$^-$.
The model bias is thus small.

The Gaussian fit provides also the RMS deviation (RMSD) of our DSM.
It is worth 14.7\,ADU rms $\approx$25\,e$^-$\,rms, \textit{i.e.} the FWHM is 34.6\,ADU $\approx$60\,e$^-$, on average from Oct.~2010 till Oct.~2012.
This is only 50\% more than the read noise (RN), which is in the range 9-11\,ADU rms for the whole period (see Sect.~\ref{sect:CameraElectronics} or Table~\ref{table:Camera}).
The observed RMSD accounts for possible algorithmic inaccuracies, and certainly more predominantly, for phenomena that cannot be modeled, such as the RTS noise of hot pixels in the memory zone.

A number of outliers depart from the Gaussian behavior and resulting shoulders are visible in Fig.~\ref{fig:DarkCorrectionEvolution}-b.
By subtracting the above Gaussian fits from the histograms, we estimate the outliers to represent $\sim$5\% of the pixels in late 2010 and $\sim$20\% in late 2012.
These percentages cover all CRHs and hot pixels having non Poissonian behaviors. 
They cannot be estimated by a DSM that is based on the assumption of a sufficiently prolonged constancy.
The error can be positive or negative.
It reaches several thousands ADU, but this level concerns a minute fraction of the pixels.
The solution is to flag the hot IZ pixels, so as to exclude them from the subsequent scientific exploitation.

Note that columns containing hot MZ pixels cannot be altogether excluded since this would then be the case of every column.
We must consequently rely on the fact that their relatively unpredictable dark contributions should compensate mutually.
However, as mentioned, this leads to a noise that probably dominates the RMSD.

\subsection{Evolution of the dark current distribution in IZ} \label{sect:IZ_distribution} 

\begin{figure}[h]
\includegraphics[width=\linewidth]{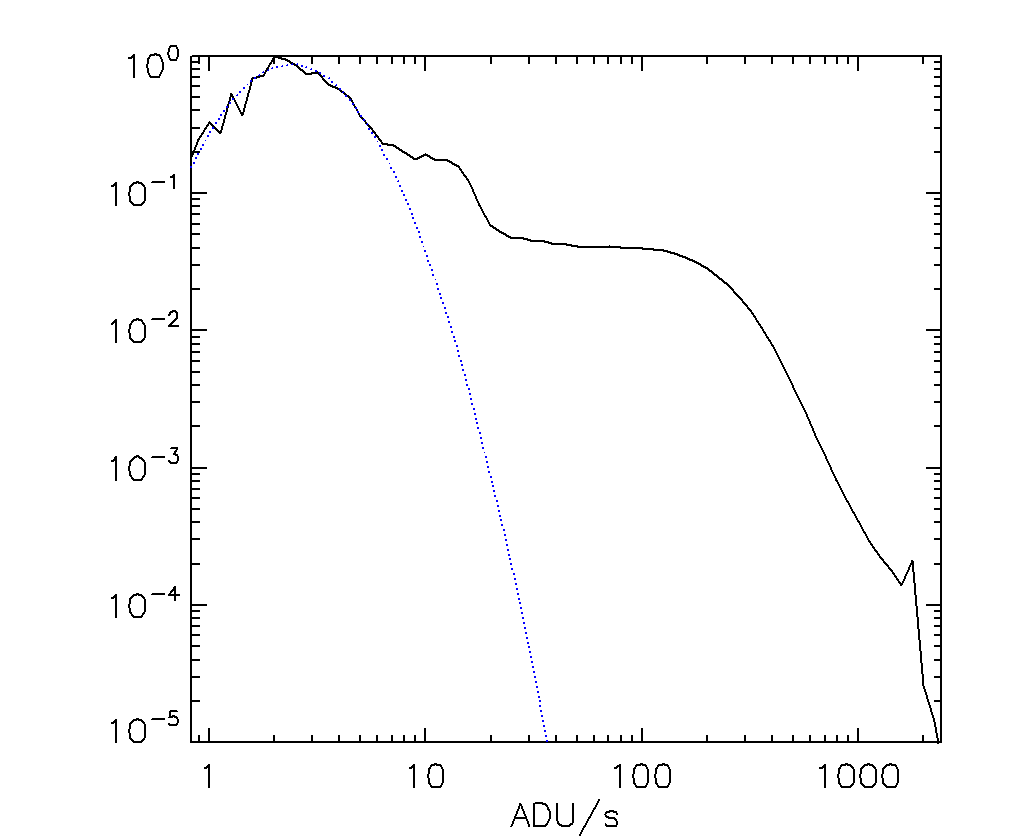}
\caption{
Normalized histogram of the Jul.~2010 -- Nov.~2012 maps reconstructed by the dark signal model (DSM) for the image zone.
The histogram mode represents the cool pixels.
It can be fitted by a lognormal function (blue dotted line), which gives 2.39\,ADU$\cdot\mathrm{pxl}^{-1}\!\cdot\mathrm{s}^{-1}$ $\approx$ 4.1\,e$^-\!\cdot\mathrm{pxl}^{-1}\!\cdot\mathrm{s}^{-1}$.
The first small shoulder around 6-20\,ADU$\cdot\mathrm{pxl}^{-1}\!\cdot\mathrm{s}^{-1}$ comprises the pixels that are polluted by some spurious signal in the bottom of the image.
The second shoulder describes the hot pixels dark current distribution.
It appears flat from 10--20\,ADU$\cdot\mathrm{pxl}^{-1}\!\cdot\mathrm{s}^{-1}$ up to 200\,ADU$\cdot\mathrm{pxl}^{-1}\!\cdot\mathrm{s}^{-1}$.
The hottest pixels deliver 2000\,ADU$\cdot\mathrm{pxl}^{-1}\!\cdot\mathrm{s}^{-1}$ $\approx$ 3.5\,ke$^-\!\cdot\mathrm{pxl}^{-1}\!\cdot\mathrm{s}^{-1}$.
}
\label{fig:ZI_Histogram} 
\end{figure}

The normalized histogram of all DSM maps for IZ is plotted in Fig.~\ref{fig:ZI_Histogram}.
Its peak corresponds to the cool (non-hot) pixels.
Fitting a lognormal distribution \citep{BaerR.L.2006}, and looking at its mode gives 2.39\,ADU$\cdot\mathrm{pxl}^{-1}\!\cdot\mathrm{s}^{-1}$ for $CPDC$, the cool pixel dark current.
Using Eq.~\eqref{eq:G15}, $CPDC$ = 2.39\,ADU$\cdot\mathrm{pxl}^{-1}\!\cdot\mathrm{s}^{-1}$ $\times$ 1.70\,e$^-$/ADU = 4.1\,e$^-\!\cdot\mathrm{pxl}^{-1}\!\cdot\mathrm{s}^{-1}$.
This is very close to the value estimated in Sect.~\ref{sect:APrioriDC}, which was 3.6 or 4.4\,e$^-\!\cdot\mathrm{pxl}^{-1}\!\cdot\mathrm{s}^{-1}$ at $-7.2$\degr C, depending on the physical modeling assumption.
We remark that the persistent solar `imprint' visible in Fig.~\ref{fig:ZI_ZM}-a does not show up in the histogram.
It is indeed of order 0.2\,ADU$\cdot\mathrm{pxl}^{-1}\!\cdot\mathrm{s}^{-1}$ (not reported in this paper) and gets merged in the main mode of the histogram.

The first shoulder of the distribution, at about 6--20\,ADU$\cdot\mathrm{pxl}^{-1}\!\cdot\mathrm{s}^{-1}$, corresponds to a pollution by an unknown signal near the image bottom, as already pointed out in the bottom of Fig.~\ref{fig:ZI_ZM}-a.

The second shoulder is flat and extends from 30\,ADU$\cdot\mathrm{pxl}^{-1}\!\cdot\mathrm{s}^{-1}$ up to $\sim$2,000\,ADU$\cdot\mathrm{pxl}^{-1}\!\cdot\mathrm{s}^{-1}$ $\approx$ 3,400\,e$^-\!\cdot\mathrm{pxl}^{-1}\!\cdot\mathrm{s}^{-1}$.
This plateau represents the detectable hot pixels, which dark current values is quasi equiprobable over at least one order of magnitude.
This study allows defining a threshold at 30\,ADU$\cdot\mathrm{pxl}^{-1}\!\cdot\mathrm{s}^{-1}$, \textit{viz.} $\sim$50\,e$^-\!\cdot\mathrm{pxl}^{-1}\!\cdot\mathrm{s}^{-1}$, for flagging the hot pixels in the image zone.
Indeed, this value belongs to the second shoulder and not to the first one. 
Thus, the 30\,ADU$\cdot\mathrm{pxl}^{-1}\!\cdot\mathrm{s}^{-1}$ threshold identifies hot pixels with no risk to erroneously trigger in the bottom part of the image that is contaminated by signal in excess.
As to the hottest pixels, they have a hundred times less dark current than specified by the manufacturer (3,400\,e$^-\!\cdot\mathrm{pxl}^{-1}\!\cdot\mathrm{s}^{-1}$ instead of 300\,ke$^-\!\cdot\mathrm{pxl}^{-1}\!\cdot\mathrm{s}^{-1}$, see Sect.~\ref{sect:APrioriDC}).

\begin{figure}[h!]
\includegraphics[width=\linewidth]{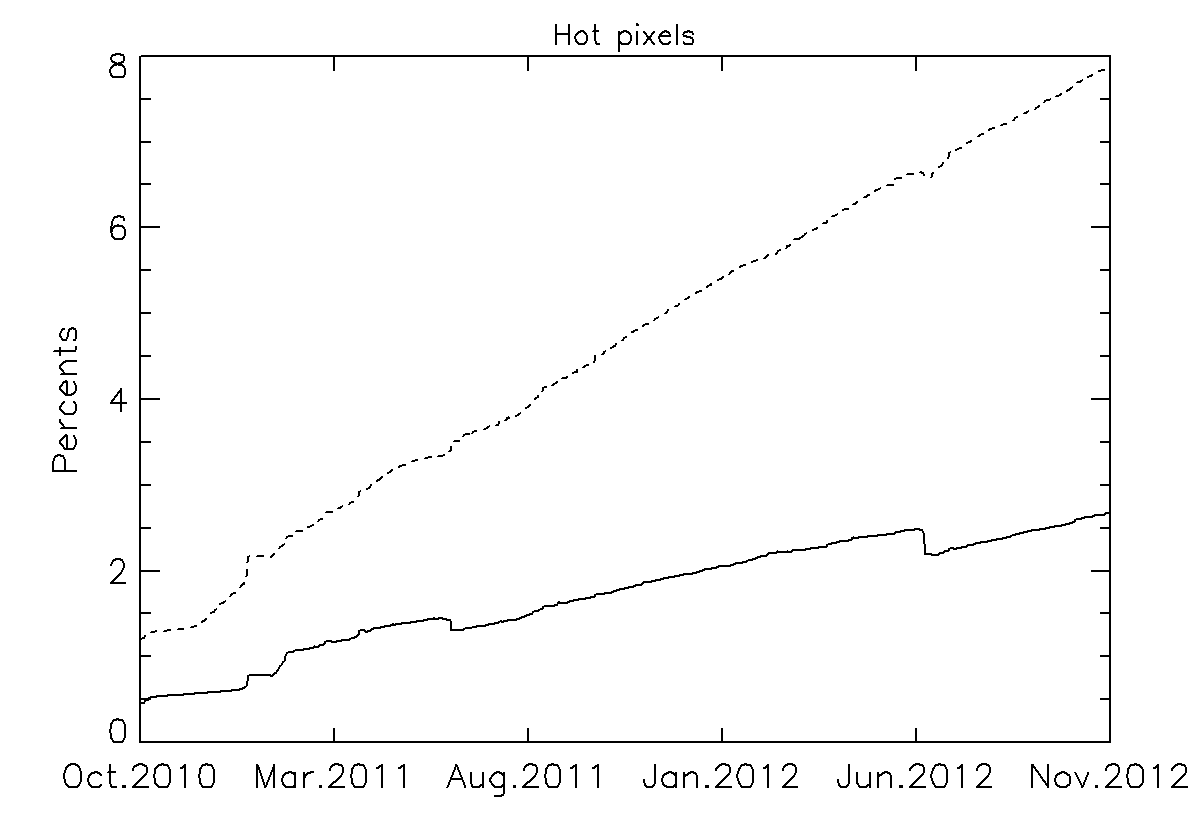}
\caption{
Temporal evolution of the fractional area covered by hot pixels with respect to the whole image zone. 
The plain line represents hot pixels delivering more than 250\,e$^-\!\cdot\mathrm{pxl}^{-1}\!\cdot\mathrm{s}^{-1}$, 
while the dash line corresponds to warm pixels delivering from 50 to 250\,e$^-\!\cdot\mathrm{pxl}^{-1}\!\cdot\mathrm{s}^{-1}$.
Note the two sudden drops in the otherwise gradual growth of the number of hotter pixels.
They occur in mid June 2011 and mid June 2012, when 2-3 day bakeouts happened.
}
\label{fig:ZI_Evolution} 
\end{figure}

We now investigate the temporal evolution of the hot pixels.
We split their severity in two separate ranges. 
The lower range goes from the limit of hot pixel detectivity (50 e$^-\!\cdot\mathrm{pxl}^{-1}\!\cdot\mathrm{s}^{-1}$) up to $\theta=250$\,e$^-\!\cdot\mathrm{pxl}^{-1}\!\cdot\mathrm{s}^{-1}$.
The upper range contains the hotter pixels, delivering more dark current than $\theta$.
The value of $\theta$ has been chosen to evidence the annealing threshold effect discussed below.
The evolution of both percentages is plotted in Fig.~\ref{fig:ZI_Evolution}.

For the lower range, the rate of hot pixel ignition is slightly decreasing over the mission lifetime, and worth $\sim$350 new hot pixels per day, \textit{viz.} $\sim$3\% of the CCD image zone per year.
For the higher range, it appears constant and worth $\sim$150 new hot pixels per day, \textit{viz.} $\sim$1.2\% of the CCD image zone per year.
This makes a total of $\sim$500 new hot pixels per day, which is very comparable to the 1000 new hot pixels per day measured for a CCD having twice SODISM format by \citet{Polidan2004,Baggett2012}. 
Their HST/WFC3 UVIS experiment uses CCDs of the E2V 43 series, which format is 2048 $\times$ 4096, pixel size is 15\,$\mu$m (instead of 13.5\,$\mu$m in our case).
Like SODISM's, those CCDs have a buried channel, an MPP implant, and are backthinned.

We additionally remark that the 50\,e$^-\!\cdot\mathrm{pxl}^{-1}\!\cdot\mathrm{s}^{-1}$ criterion leads to a $\sim$10\,\% area coverage by hot pixels in late 2012, while we have found 20\,\% pixels departing from the Gaussian behavior in the previous section.
This is because many of those fall below our threshold (see Fig.~\ref{fig:ZI_Histogram}).
But lowering it would flag too many pixels as being `hot', among which many would moreover become false positives.

The drops in the slope observed in mid June 2011 and mid June 2012 in Fig.~\ref{fig:ZI_Evolution}, especially for the hotter pixel curve, are coincident with the two SODISM bakeout operations.
Indeed they reveal that the pixels having a dark current higher than $\theta$ were instantaneously (but only partially) annealed on those occasions.
The faster annealing of hotter pixels as compared to the annealing of the `warm' ones has been noted by \citet{Polidan2004} and studied by \citet{Marshall2005}.
An interesting possibility inferred from the latter reference is that the hotter pixels of SODISM may be to some extent self-annealing constantly due to the `lukecold' operating temperature of its CCD.

Some more verifications have been made.
The average dark current of the whole image zone is found to increase from a few e$^-\!\cdot\mathrm{pxl}^{-1}\!\cdot\mathrm{s}^{-1}$ in mid 2010 up to 30\,e$^-\!\cdot\mathrm{pxl}^{-1}\!\cdot\mathrm{s}^{-1}$ in late 2012.
This can be compared to the equivalent estimation that can be made for the memory zone.
Interestingly, the integrated dark current in the MZ is found to be consistently 20\% larger than the IZ counterpart. 
If this is real, it could mean that the shield of the MZ favors the generation of non ionizing collisions at the origin of displacement damages and hence, of hot pixels.

Finally, it is established that the mean level of the first dark row of the MZ increases gradually from $\sim$0\,e$^-$/pxl in mid 2010 up to $\sim$40\,e$^-$/pxl in late 2012.
This is possibly due to surface dark current seeping increasingly into the pixels during the frame transfer.

%%%%%%%%%%%%%%%%%%%%%%%%%%%%%%%%%%%%%%%%%%%%%%%%%%%%%%%%%%%%%%%%%%%%%%%%%%%%%%%%%%%%%%%%%%%%%%%%%%%%%%

\section{Summary and conclusions} \label{sect:Conclusions} 

Concerning the camera of the PICARD-SODISM telescope equipped with its Flight CCD\,\#60, we have learned the following:
\begin{enumerate}
      \item The inverse of the camera gain is $G_{16}^{-1} \approx\,0.85\,\mathrm{e}^-/\mathrm{ADU}_\mathrm{16\,bit}$ when the image is coded on 16 bits, 
               and $G^{-1}_{15} \approx1.70\,\mathrm{e}^-/\mathrm{ADU}_\mathrm{15\,bit}$ when it is coded on 15 bits, as is more commonly the case.
               That value has been obtained from flight data by means of the `photon transfer technique' \citep{Janesick1987a} applied to a special sequence of dark frames recorded in July 2010.
      \item The overall camera read noise (RN) is measured to be 15\,e$^-$ rms in July 2010 and 20\,e$^-$ rms  in late 2012. 
               This is respectively four and five times larger than the intrinsic CCD read noise. The camera RN appears contaminated by an orbital variability.
      \item At the operating temperature of the Flight CCD ($-7.2$\degr C), the dark current of its non hot pixels is predicted to be $CPDC$ = 3.6 or 4.4\,e$^-\!\cdot\mathrm{pxl}^{-1}\!\cdot\mathrm{s}^{-1}$ depending on the adopted physical model.
      \item The dark signal components of the CCD image zone and memory zone have been estimated daily by the dark signal model (DSM) presented in this paper. 
               Its outcome is online and exploited by SODISM Level~1 products.
      \item According to our DSM, the dark current of non hot pixels is estimated to be $CPDC$ = 4.1\,e$^-\!\cdot\mathrm{pxl}^{-1}\!\cdot\mathrm{s}^{-1}$ in the image zone, in good agreement with the expectation for a CCD at $-7.2$\degr C.
      \item According to our DSM, the hottest dark pixels produce 3.5\,ke$^-\!\cdot\mathrm{pxl}^{-1}\!\cdot\mathrm{s}^{-1}$ at $-7.2$\degr C.
      \item The DSM is shown to exhibit a global bias of $\sim$5\,e$^-$ and an rms deviation (RMSD) of 25\,e$^-$\,rms, for images taken with a 7.4\,s integration time.
      \item A threshold can be reliably set at 50\,e$^-\!\cdot\mathrm{pxl}^{-1}\!\cdot\mathrm{s}^{-1}$ to flag the hot pixels of the image zone for their subsequent (optional) dismissal by science investigations.
      \item With this threshold, $\sim$1.5\,\% of the Flight CCD pixels are declared hot at commissioning (Oct. 2010) and $\sim$11\,\% in Dec. 2012.
\end{enumerate}

In conclusion, more general observations and results are recapitulated:
\begin{enumerate}
      \item Dark signal correction is complicated when the CCD has a frame transfer architecture, and especially when its temperature is not cold enough and/or when the readout is not fast enough to essentially suppress the dark signal contribution of the memory zone.
      \item Frame transfer CCDs should therefore be employed when their advantages ($\sim$100\,\% observational duty cycle, avoidance of a mechanical shutter) surpass their inconveniences.
      \item Temperatures below $-10$\degr C look consequently desirable for frame transfer CCD operations, although such `lukecold' temperatures may mitigate CTE issues and the growth rate of hot pixels.
      \item It is nevertheless possible to model and correct for the dark signal of a frame transfer CCD by applying the method presented in this paper.
      \item Aspects of our method can be useful to other applications. Modeling the dark signal of full frame CCDs is a natural example.
               The unbalanced Haar transform of \citet{Fryzlewicz2007}, or adapted versions thereof, appear particularly powerful to fit piecewise constant functions in various situations.
      \item Onboard estimations of \textit{e.g.} the CCD offset or its read noise should not be rounded to integer values.
      \item It is a good idea to plan flight operations that cycle integration time across its admissible range, automatically and regularly, especially for the dark frames.
      \item Bakeout operations at room temperature for few days do anneal a fraction of the hot pixels, and particularly the hotter ones,
               but loosing few days of observation can be considered worthless with respect to their limited effectiveness.
               Reciprocally, many more bakeouts could be programmed so as to maintain the CCD near to its pristine state.
\end{enumerate}

%%%%%%%%%%%%%%%%%%%%%%%%%%%%%%%%%%%%%%%%%%%%%%%%%%%%%%%%%%%%%%%%%%%%%%%%%%%%%%%%%%%%%%%%%%%%%%%%%%%%%%

\begin{acknowledgements}
PICARD-SODISM is a project supported by the Centre National d'Etudes Spatiales (CNES), the CNRS/INSU, and the Belgian Space Policy (BELSPO).
The SODISM instrument has been designed and built in collaboration between CNRS-LATMOS and CNES.
It is operated in coordination by the Scientific Mission Center (CMS-P) at BUSOC, Belgium, CNES, and LATMOS.
The authors acknowledge gratefully the work of Claudio~Queirolo, Anuschka~Helderweirt, Dirk~Pauwels at BUSOC for their modification of the in-flight payload operations, 
as well as Abdenour~Irbah, Momar~Ciss\'e, Marc~Lin, and Christophe~Dufour at LATMOS for their development of the SODISM processing chain, and for their helpful support.
The authors would like to also thank Peter Maggs and David Morris from E2V, and 
Pernelle Bernardi, Didier Tiph\`ene, Vincent Lapeyrere, and Jean Tristan~Buey from LESIA for their answers to our questions.
J.-F.\,H. thanks Francis Dalaudier for useful suggestions.
J.-F.\,H. has also appreciated the help of Caroline\,Guerin, Andr\'e-Jean\,Vieau, Yann\,Delcambre, and Olivier\,Thauvin of LATMOS for their kind support to the work that is reported here.
\end{acknowledgements}

\bibliographystyle{aa} % style aa.bst
\bibliography{SODISM_DarkSignal_JFH} % your references Yourfile.bib

\end{document}